%% file: mmWave.tex
\documentclass{IEEEtran}

\pdfoutput=1

\usepackage{amsmath}
\usepackage{amssymb}
\usepackage{mathtools}
\usepackage{mathdots}
\usepackage[caption=false]{subfig}
\usepackage{graphicx}
\usepackage{epstopdf}
\usepackage{cite}
\usepackage{relsize}
\usepackage{mathrsfs}
\usepackage{algorithm,algorithmic}
\usepackage{multirow}
\usepackage{color}
\usepackage{enumitem}
\usepackage{soul}
\usepackage{hyperref}

\DeclareMathOperator{\an}{a}

\DeclareMathOperator{\ap}{ap}
\DeclareMathOperator{\di}{d}

\DeclareMathOperator{\dl}{dl}

\DeclareMathOperator{\uplink}{ul}
\DeclareMathOperator{\ue}{sta}

\DeclareMathOperator{\rank}{rank}
\DeclareMathOperator{\snr}{SNR}

\DeclareMathOperator{\eq}{eq}

\DeclareMathOperator{\inter}{int}

\DeclareMathOperator{\rf}{rf}

\DeclareMathOperator{\tx}{t}

\DeclareMathOperator{\sub}{sub}

\newcommand{\E}{\mathbb{E}}
\newcommand{\C}{\mathbb{C}}

\newcommand{\Bset}{\mathcal{B}}

\newcommand{\Pset}{\mathcal{P}}

\newcommand{\Gset}{\mathcal{G}}

\newcommand{\Kset}{\mathcal{K}}

\newcommand{\cgaussian}{\mathcal{CN}}

\newcommand{\HH}{\mathbf{H}}
\newcommand{\PP}{\mathbf{P}}
\newcommand{\SSS}{\mathbf{S}}

\newcommand{\A}{\mathbf{A}}
\newcommand{\B}{\mathbf{B}}

\newcommand{\I}{\mathbf{I}}

\newcommand{\RR}{\mathbf{R}}

\newcommand{\aaa}{\mathbf{a}}
\newcommand{\bb}{\mathbf{b}}

\newcommand{\h}{\mathbf{h}}
\newcommand{\g}{\mathbf{g}}
\newcommand{\p}{\mathbf{p}}

\newcommand{\s}{\mathbf{s}}

\newcommand{\vv}{\mathbf{v}}

\newcommand{\x}{\mathbf{x}}
\newcommand{\y}{\mathbf{y}}
\newcommand{\z}{\mathbf{z}}

\newcommand{\vh}{\hat{v}}
\newcommand{\hbh}{\hat{\mathbf{h}}}
\newcommand{\vbh}{\hat{\mathbf{v}}}
\newcommand{\wh}{\hat{w}}
\newcommand{\wbh}{\hat{\mathbf{w}}}

\newcommand{\ze}{\mathbf{0}}
\newcommand{\one}{\mathbf{1}}

\begin{document}

\title{Beamforming Algorithm for Multiuser Wideband Millimeter-Wave Systems with Hybrid and Subarray Architectures}

\author{Carlos~A.~Viteri-Mera,~\IEEEmembership{Senior~Member,~IEEE},~and~Fernando~L.~Teixeira, \IEEEmembership{Fellow,~IEEE}% <-this % stops a space
\thanks{C. Viteri-Mera is with the Department of Electronics Engineering, Universidad de Nari\~no, Pasto, Colombia, email: \href{mailto:caviteri@udenar.edu.co}{caviteri@udenar.edu.co}.}% <-this % stops a space
\thanks{F. Teixeira is with the ElectroScience Laboratory, The Ohio State University, 1330 Kinnear Rd., Columbus, OH, 43212 USA, email: \href{mailto:teixeira.5@osu.edu}{teixeira.5@osu.edu}.}% <-this % stops a space
\thanks{Manuscript updated on March 15, 2019.}}

\maketitle

\begin{abstract}
We present a beamforming algorithm for multiuser wideband millimeter wave (mmWave) communication systems where one access point uses hybrid analog/digital beamforming while multiple user stations have phased-arrays with a single RF chain. The algorithm operates in a more general mode than others available in literature and has lower computational complexity and training overhead. Throughout the paper, we describe: \emph{i}) the construction of novel beamformer sets (codebooks) with wide \emph{sector} beams and \emph{narrow} beams based on the orthogonality property of beamformer vectors, \emph{ii}) a beamforming algorithm that uses training transmissions over the codebooks to select the beamformers that maximize the received sum-power along the bandwidth, and \emph{iii}) a numerical validation of the algorithm in standard indoor scenarios for mmWave WLANs using channels obtained with both statistical and ray-tracing models. Our algorithm is designed to serve multiple users in a wideband OFDM system and does not require channel matrix knowledge or a {particular channel structure.} Moreover, we {incorporate antenna-specific aspects in the analysis, such as antenna coupling, element radiation pattern, and beam squint. Although there are no other solutions for the general system studied in this paper}, we characterize the algorithm's achievable rate and show that it attains more than 70\% of the spectral efficiency (between 1.5 and 3 dB $\snr$ loss) with respect to ideal fully-digital beamforming in the analyzed scenarios. We also show that our algorithm has similar sum-rate performance as other solutions in the literature for some special cases, while providing significantly lower computational complexity (with a linear dependence on the number of antennas) and shorter training overhead.
\end{abstract}


\input{Introduction}

\input{SystemModel}

\input{Problem}

\input{Codebook}

\input{Algorithm}

\input{Results}

\input{Conclusions}

%\IEEEtriggeratref{39}
\bibliographystyle{IEEEtran}
\bibliography{mmWave}

\begin{IEEEbiography}{Carlos A. Viteri-Mera} (SM'18)
received the B.S. degree in Electronics and Telecommunications Engineering from Universidad del Cauca (Colombia) in 2006, the M.S. degree in Electronics and Computer Engineering from Universidad de los Andes (Colombia) in 2010, and the Ph.D. degree in Electrical and Computer Engineering from The Ohio State University (Columbus, OH) in 2018. From 2013 to 2016 he was a Fulbright Scholar during his doctoral studies. Dr. Viteri is currently an Assistant Professor at the Universidad de Nari\~no, Pasto, Colombia. His research focuses on physical layer aspects of wireless communications.
\end{IEEEbiography}

\begin{IEEEbiography}{Fernando L. Teixeira} (F'15)
received the B.S. and M.S. from the Pontifical Catholic University of Rio de Janeiro, Brazil in 1991 and 1995 respectively, and Ph.D. degree from the University of Illinois at Urbana-Champaign, USA in 1999, all in electrical engineering. From 1999 to 2000, he was a Postdoctoral Associate with the Massachusetts Institute of Technology. In 2000, he joined the The Ohio State University, where he is now a Professor with the Department of Electrical and Computer Engineering and also affiliated with the ElectroScience Laboratory.
 
Dr. Teixeira served in the past as Chair of the Joint IEEE AP/MTT-S Columbus Chapter and was an Associate Editor and Guest Editor for the IEEE ANTENNAS AND WIRELESS PROPAGATION LETTERS. He currently serves as Associate Editor for {\it IET Microwaves, Antennas, and Propagation}. He is the recipient of the CAREER Award from the National Science Foundation, the triennial Booker Fellowship from the International Union of Radio Science (USNC/URSI), and the Outstanding Young Engineer Award from the IEEE Microwave Theory and Techniques Society (MTT-S).
\end{IEEEbiography}

\end{document}

%% file: Introduction.tex
\section{Introduction}
\label{sec_intro}
\IEEEPARstart{M}{illimeter} wave (mmWave) systems are emerging technologies for future wireless communication networks. These systems use carrier frequencies around 28, 38, 60, and 72 GHz, where large bandwidths are available to alleviate spectrum scarcity affecting current cellular networks~\cite{boccardi2014}. Propagation at mmWave frequencies is characterized by large free-space losses and strong atmospheric attenuation~\cite{levis2010}. {These features are favorable for short-range ($<$100 meters) communications since large propagation losses naturally decrease interference across dense small-cells and can be compensated to a certain extent using high-gain antennas. Uniform linear arrays (ULAs)}~\cite{lota2017} {or planar arrays}~\cite{hong2017} {are typical high-gain antenna implementations at mmWave which can be mounted in compact terminals given the small wavelength}~\cite{rappaport2014}. {However, these antenna arrays have narrow pencil-like radiation patterns that require beamforming techniques to align their main beams according to the specific scenario.}

%A number of works have explored the propagation characteristics of mmWave frequencies and their application to 5G cellular networks (e.g., see~\cite{rappaport2017} for an overview of this topic). In particular, 60 GHz bands have been studied for indoor wireless communications since atmospheric absorption losses in these bands are too strong for longer (outdoor) links~\cite{rappaport2014}. Current and developing standards for indoor WLAN use carriers around 60 GHz, where the available unlicensed bandwidth is approximately 5 GHz~\cite{iso2014,maltsev2017}. Applications such as ultra-high-definition video streaming, augmented and virtual reality, mass data distribution, {and indoor wireless backhauling will be enabled by this bandwidth}~\cite{maltsev2017}.

Despite the advantages of mmWave systems, many challenges remain open to fully exploit their potential. {One of the most active research areas in mmWave is the design of beamforming procedures that take into account the specific hardware constraints at these frequencies}~\cite{roh2014}. {For example, fully-digital beamforming, which requires one RF chain for every antenna element, is unfeasible in mmWave due to the large number of small antenna elements deployed in typical arrays.} For this reason, several hybrid analog/digital beamforming architectures (which use phase shifters, power splitters/combiners, and switches in the RF domain together with digital baseband processing) have been proposed as practical solutions for mmWave terminals~\cite{ayach2014,alkhateeb2014,alkhateeb2016,mendez2016}. The hybrid beamforming design problem, which is the focus of this paper, {refers to the selection of phase-shifter configurations and digital processing matrices that maximize a given performance criteria under given hardware constraints.} In this work, we propose a beamforming algorithm for mmWave systems with the following features:
\begin{enumerate}[label=F\arabic*.]
\item One single access point (AP) serving multiple user stations (STAs).
\item Fully-connected hybrid architecture at the AP, where a few RF chains are connected to a large antenna array through a network of phase shifters and power splitters/combiners.
\item Subarray architecture at the STAs, where a single RF chain can use the full antenna array or smaller subarrays selected with an RF switch.
\item Wideband (OFDM) operation.
\item Blind operation: it does not require knowledge (nor estimation) of the channel matrix.
\item A beamforming procedure based on hierarchichal codebooks.
\end{enumerate}

\subsection{Related Work}
The hybrid beamforming problem for single-user narrowband mmWave systems was first studied in \cite{ayach2014} and \cite{alkhateeb2014}. In \cite{ayach2014}, assuming perfect channel state information (CSI), {the authors design hybrid transmit beamformers to approximate the linear precoder that maximizes the spectral efficiency.} This is done by leveraging the spatially sparse nature of the mmWave channel and considering phase shifters with quantized states. This work is expanded in \cite{alkhateeb2014}, where channel estimation is formulated as a sparse reconstruction problem. This approach uses sets of hybrid beamforming configurations, called hierarchical beamforming codebooks, that provide main beams of different widths. These codebooks are used to perform training transmissions over increasingly narrow angular regions in space, converging to the sharpest beam that maximizes the received power. {Both} \cite{ayach2014} and \cite{alkhateeb2014} {focus on systems having features F2 and F6 only.}

\color{black}
The use of hierarchical beamforming codebooks (feature F6) has become a design principle for beamforming {in different mmWave applications.} For example, the IEEE 802.11ad 60 GHz WLAN standard uses beam training over a hierarchical codebook \cite{wang2009,iso2014,hosoya2015}. {Reference} \cite{song2017} {presents a codebook construction methodology for hybrid beamforming based on planar antenna arrays in single-user narrowband systems.} In the analog beamforming case, \cite{hur2013} {presents a hierarchical beamforming procedure that accounts for wind-induced antenna movement in mmWave links.

Another widely used design criteria for hybrid beamformers in mmWave systems is the spectral efficiency (mutual information) maximization.} In the single-user frequency-selective case (feature F4), \cite{alkhateeb2016} proposes a greedy optimization technique to maximize mutual information under perfect CSI. Reference \cite{park2017} obtains closed-form solutions to a relaxed mutual information maximization problem and examine subarray structures for single-user frequency-selective hybrid beamforming (features F2-F4) assuming perfect knowledge of the channel sample covariance matrix.

Reference \cite{raghavan2017} demonstrates that multiuser hybrid beamforming (features F1 and F2) is better suited for networks where users are well-separated in the angular domain. A multiuser hybrid beamforming design solution for narrowband mmWave is presented in \cite{alkhateeb2015}, where the analog beamformer is obtained by training over a hierarchical codebook and the digital beamformer is calculated in a conventional zero-forcing approach (features F1, F2, and F6).

For multiuser frequency-selective hybrid beamforming (features F1, F2, and F4), \cite{yu2016} formulates the design problem such that the hybrid beamformer approximates the fully digital solution found with block-diagonalization (BD) \cite{spencer2004}. The work in \cite{yu2016} assumes perfect CSI with a total power constraint and analog beamforming with infinite resolution phase shifters. Reference \cite{sohrabi2017} also investigates hybrid beamforming in the multiuser frequency-selective case assuming perfect CSI and infinite resolution phase shifters; its approach is to simplify the problem to a frequency-flat equivalent by using an average of the channel matrices along subcarriers. 

{Note that most of the works referenced above require additional algorithms to estimate the wideband channel matrix, thus increasing their complexity and training overhead.} Reference \cite{venugopal2017} presents several methods for wideband channel estimation in the time and frequency domains.

Antenna-specific aspects of wideband mmWave beamforming have also been studied recently, with focus on the array response's frequency dependence (\emph{beam squint} effect). References \cite{cai2016} and \cite{cai2017} {study the capacity loss in mmWave systems due to beam squint and propose a beamforming design to compensate for this effect.} %Other antenna aspects of beamforming for mmWave terminals are addressed in \cite{hong2017}.

\subsection{Contributions}

{This paper presents a new beamforming algorithm incorporating two main contributions with respect to previous works: }
\begin{enumerate}
\item {The algorithm operates in a more general setting than other beamforming solutions (including all the 6 features above). Previous algorithms work in systems that can be considered special cases of this general setting (for example, single-user or narrowband systems).}
\item {The algorithm reduces the computational complexity and training overhead with respect to previous works by leveraging the hybrid and subarray hardware architectures. More specifically, our algorithm's computational complexity and training overhead increases linearly with the number of antennas, while previous solutions have a quadratic dependence.}
\end{enumerate}

{To achieve these two objectives, the algorithm uses novel analog hierarchical beamforming codebooks specifically designed for hybrid and subarray configurations. The codebooks provide \emph{sector} beams and \emph{narrow} beams to allow training transmissions over increasingly narrow regions in the angular domain. The codebooks are based on the orthogonality of beamforming vectors in uniform antenna arrays, which allows an exact number of narrow beams to overlap with a given sector beam regardless of the frequency. Thus, they are well-suited to compensate for beam squint effects.}

{The algorithm decouples the design process into two stages: the analog beam selection procedure and the digital beamforming. First, in the beam selection procedure, the algorithm obtains analog beamformers that maximize the estimated sum-power across subcarriers for each user. The beam selection uses the orthogonal beamforming codebooks for uplink and downlink training transmissions. In the second part, the AP calculates a BD digital beamformer that maps the user streams to the available RF chains. Our algorithm does not require channel matrix knowledge, or any particular channel structure. The only assumption about the channel is that it is reciprocal and quasi-static. Thus, the algorithm is suitable for time division duplex systems.} {We also present wideband mmWave system models that take into account beam squint, element radiation patterns, and antenna coupling effects, which are traditionally ignored in literature. }

{The algorithm's performance is evaluated in terms of achievable sum-rate in the downlink under two channel models: \emph{i}) a statistical channel model with a discrete number of paths and uniformly distributed angles of departure and arrival, and \emph{ii}) channel matrices obtained with ray-tracing simulations from a realistic scenario taken from the IEEE 802.11ay channel models} \cite{maltsev2017}. These simulations include diffuse scattering at surfaces to model mmWave propagation characteristics. {We compare our algorithm versus fully-digital BD precoding (with one RF chain per antenna), which is the linear beamforming technique that best approximates the channel capacity }\cite{lee2006}. {We show that our algorithm's achievable sum-rate performance has only approximately a 3 dB loss with respect to the fully-digital BD solution. Given that (to the best of our knowledge) there is presently no method that operates in a system with all 6 features above, we compare our algorithm with others in current literature for specific cases (i.e. single-user, or multi-user with single-antenna terminals). We show that our algorithm provides advantages with respect to computational complexity (linear vs. quadratic dependence on the number of antennas) and training overhead in those specific cases. In some cases, this computational complexity advantages come at a cost of a marginal performance loss (0 to 3 dB SNR), while in other cases is even complemented with a better performance.}

Throughout the paper, we show application examples for ULAs and parameters taken from the IEEE 802.11ad WLAN standard. However, the principles discussed here are also applicable to 2D uniform planar arrays and other types of mmWave networks. {Note that our algorithm can be easily modified to suit specific system requirements (special cases of the 6 features).}

%% file: SystemModel.tex
\section{System Model}
\label{sec_sytemmodel}

In this section\footnote{\emph{Notation}: $a$ and $A$ are scalars, $\aaa$ is a vector, and $\A$ is a matrix. Vector and matrix sizes are defined explicitly for every variable. $(\cdot)^T$, $(\cdot)^*$, $(\cdot)^H$, and $\Vert \cdot \Vert_F$ represent transpose, complex conjugate, conjugate transpose, and Frobenius norm of a matrix, respectively. $[\A]_{m,n}$ is the element in the $m$-th row and $n$-th column of $\A$. $[\A]_{:,n}$ is the $n$-th column of $\A$. $\Vert \aaa \Vert_2 = \sqrt{\aaa^H\aaa}$ is the $\ell_2$ norm of $\aaa$. $\I_N$ is the $N \times N$ identity matrix. $\E\{\cdot\}$ denotes expected value and $\cgaussian (\ze, \RR )$ is the zero-mean circularly-symmetric complex Gaussian distribution with covariance matrix $\RR$.}, we introduce wideband (OFDM) mmWave system models for the two stages of our algorithm. We begin by defining the hardware configurations at the AP and the STAs, and then we present baseband signal models for two stages of hybrid beamforming design. In the \emph{beam selection} procedure, the algorithm searches for analog beamformers that maximize the estimated received power for each STA, alternating between uplink and downlink training transmissions. We define single-user uplink and downlink OFDM system models for this stage. In the \emph{digital beamforming} stage, the algorithm uses the analog beamformers found in the first stage to calculate a digital beamformer that maps the transmitted signals (directed to all users) to the available RF chains. Hence, we define an OFDM multiuser MIMO downlink model with hybrid beamforming. Importantly, the models presented here include wideband effects relevant in mmWave systems that are traditionally ignored in literature. For example, we include detailed frequency-dependent array response vectors that account for antenna coupling, element radiation patterns, and beam squint effects.

\subsection{Hardware Configuration}
\begin{figure}
\centering
\includegraphics[width=0.7\columnwidth]{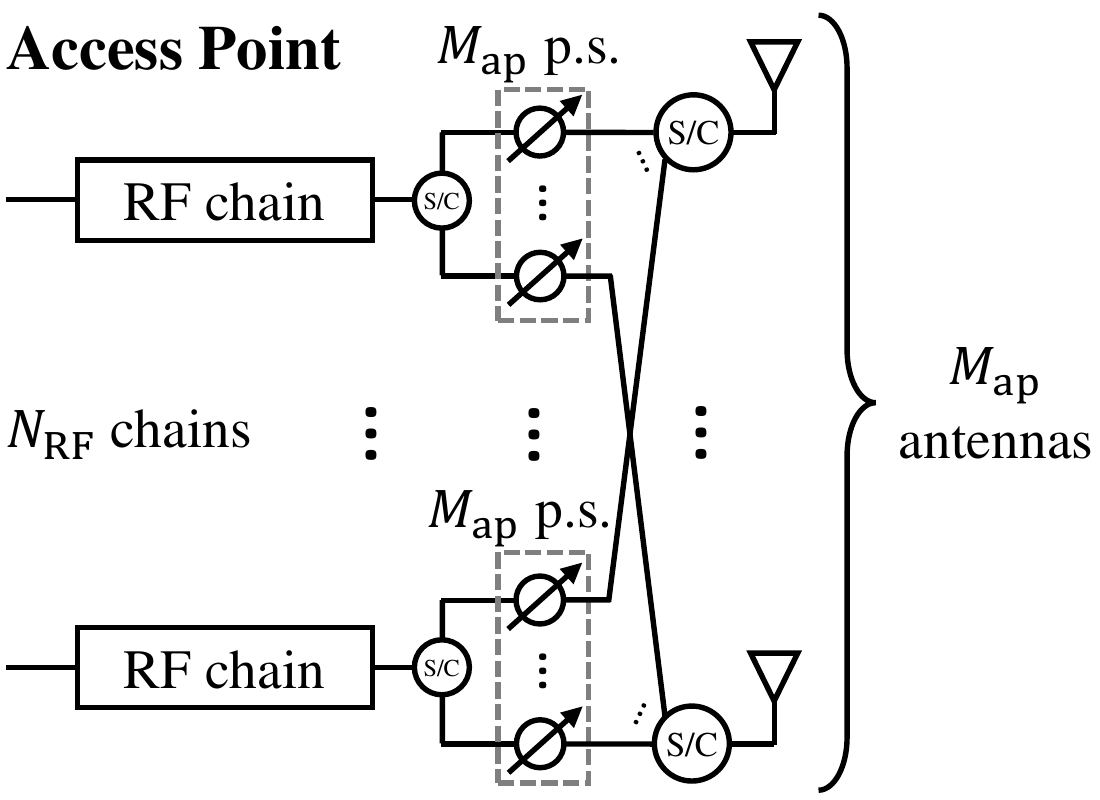}
\caption{AP hardware configuration. Each RF chain is connected to a group of $M_{\ap}$ phase shifters (p.s.) through one splitter/combiner (S/C). One phase shifter from each RF chain is connected to every antenna through other S/C.}
\label{fig_aphardware}
\end{figure}

\begin{figure}
\centering
\includegraphics[width=0.7\columnwidth]{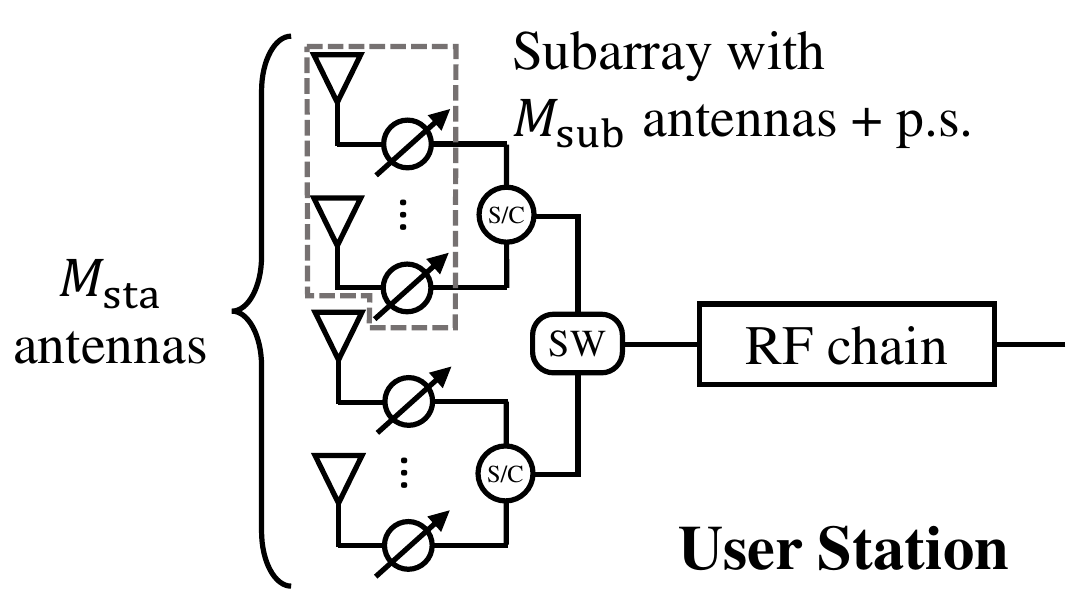}
\caption{User station configuration. Each one of the $M_{\ue}$ antennas is connected to a single phase shifter (p.s.). An RF switch allows the antenna reconfiguration to use the full array with $M_{\ue}$ antennas or a subarray with $M_{\sub}$ antennas.}
\label{fig_uehardware}
\end{figure}

We consider a mmWave system consisting of one AP and $U$ user stations (STAs). The AP has $M_{\ap}$ antennas and $N_{\rf}$ RF chains, which are fully connected as depicted in Fig. \ref{fig_aphardware}. Since the AP has multiple RF chains, it can also perform digital processing in addition to the analog beamforming, an operation known as hybrid beamforming or precoding. In the STA side, where there are usually very strict size and power constraints, we use a less demanding hardware configuration with $M_{\ue}$ antennas and a single RF chain, as shown in Fig. \ref{fig_uehardware}. This is a conventional analog phased-array configuration, with one phase shifter per antenna. However, we propose to employ a switching network so that the STA can also use a smaller subarray of $M_{\sub}$ antennas to provide wider beams at the expense of less beamforming gain. {The STAs can use either full-array or subarray configurations by controlling the RF switching network denoted by SW in Fig.} \ref{fig_uehardware}. {We assume that all phase shifters have digital control.} Therefore, there is a discrete (quantized) set of phases and beamformers available at the terminals. In Section \ref{sec_codebook}, we describe how to leverage this hardware configuration to construct discrete beamforming codebooks with both wide {sector} beams and {narrow} beams.

%\Figure[t!](topskip=0pt, botskip=0pt, midskip=0pt)[width=0.7\columnwidth]{Figures/AP.pdf}
%{AP hardware configuration. Each RF chain is connected to a group of $M_{\ap}$ phase shifters (p.s.) through one splitter/combiner (S/C). One phase shifter from each RF chain is connected to every antenna through other S/C.\label{fig_aphardware}}
%\Figure[t!](topskip=0pt, botskip=0pt, midskip=0pt)[width=0.7\columnwidth]{Figures/User.pdf}
%{User station configuration. Each one of the $M_{\ue}$ antennas is connected to a single phase shifter (p.s.). An RF switch allows the antenna reconfiguration to use the full array with $M_{\ue}$ antennas or a subarray with $M_{\sub}$ antennas.\label{fig_uehardware}}
%\begin{figure}
%\centering
%\includegraphics[width=0.7\columnwidth]{Figures/User.pdf}
%\caption{User station configuration. Each one of the $M_{\ue}$ antennas is connected to a single phase shifter (p.s.). An RF switch allows the antenna reconfiguration to use the full array with $M_{\ue}$ antennas or a subarray with $M_{\sub}$ antennas.}
%\label{fig_uehardware}
%\end{figure}
% This configuration requires $M_{\ap}N_{\rf}$ phase shifters (p.s.), $N_{\rf}$ splitters/combiners (S/C) with $M_{\ap}$ ports, and $M_{\ap}$ S/C with $N_{\rf}$ ports.

\subsection{Single-User mmWave Uplink - Beam Selection}
\label{sec_systemmodeltrainingdownlink}
\begin{figure}
\centering
\subfloat[]{\includegraphics[width=\columnwidth]{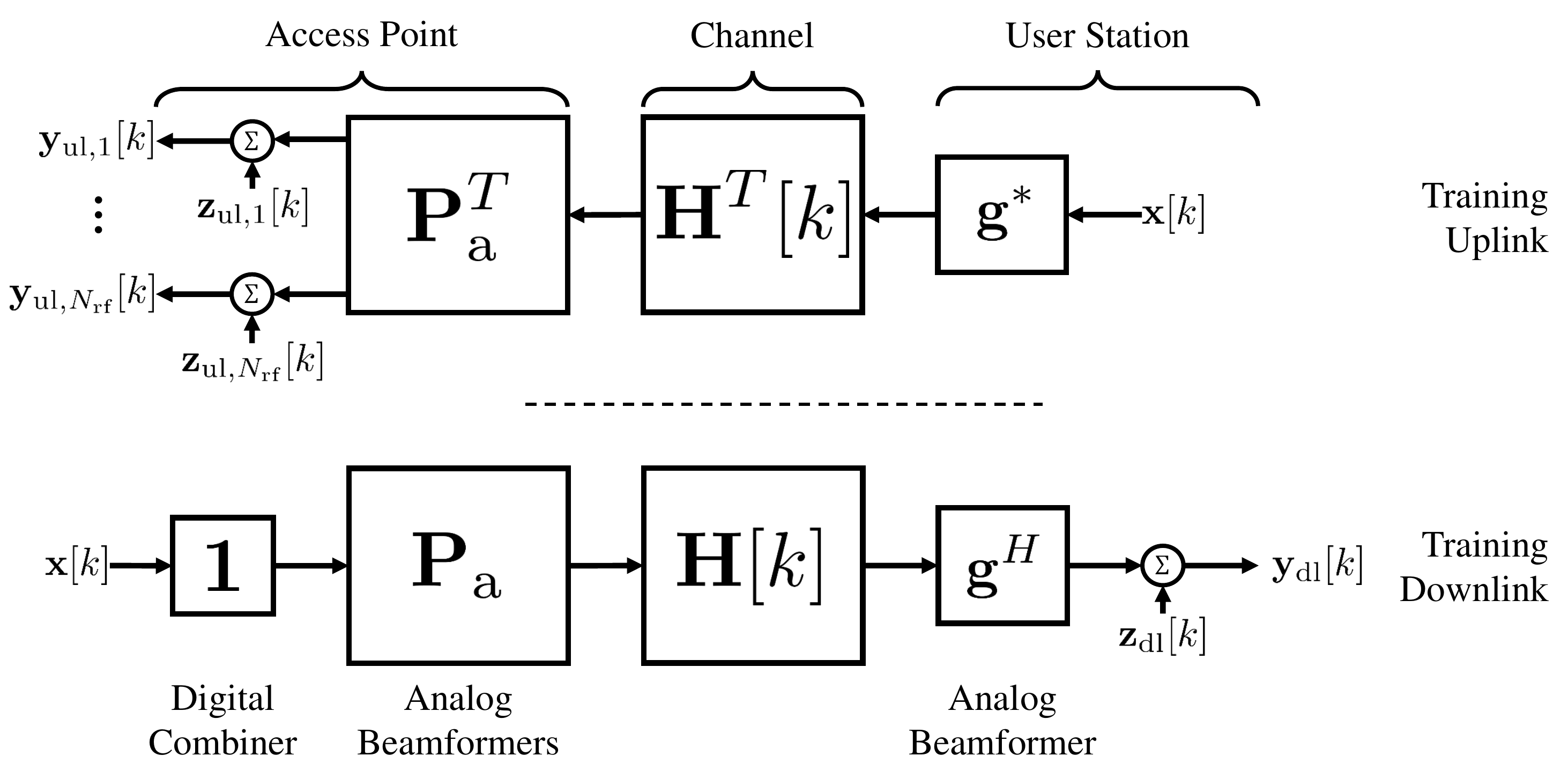} \label{fig_modeltraining}}\\
\subfloat[]{\includegraphics[width=\columnwidth]{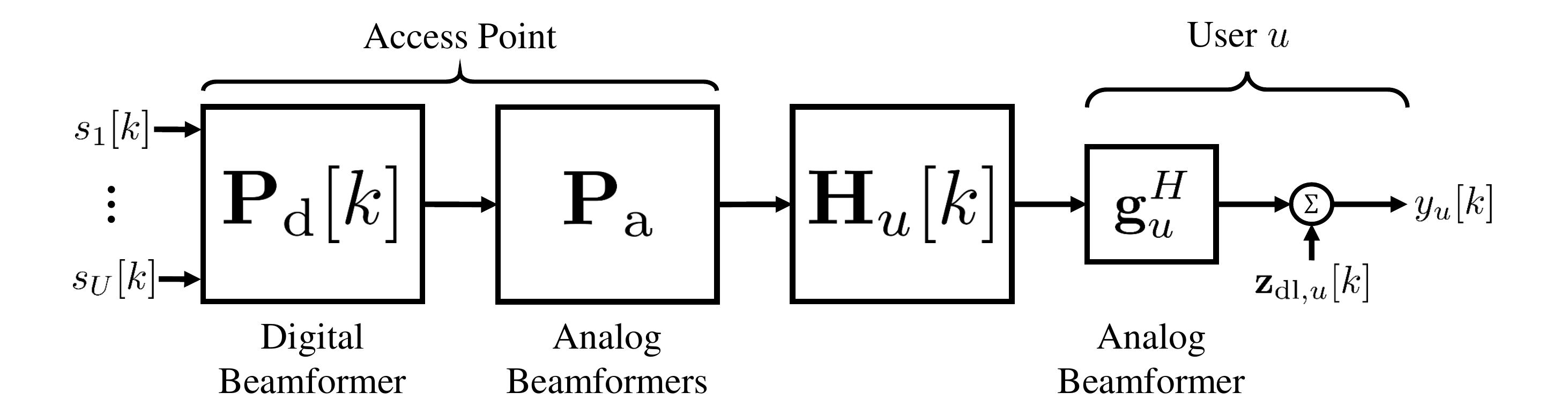}\label{fig_modelmultiuser}}
\caption{System model. (a) Training uplink and downlink, and (b) Downlink for user $u$.}
\label{fig_model}
\end{figure}

Consider a wideband wireless communication system represented in complex baseband. The system uses an OFDM waveform with $K$ subcarriers, and we assume that the OFDM cyclic-prefix removes all of the inter-symbol interference. In the \emph{beam selection} procedure, the STA transmit a training sequence $\x[k] \in \C^{1 \times T}$ in the uplink, which spans $T$ time-domain OFDM symbols (each row represents different time samples). The received uplink vector signal, which represents the output of all RF chains at the AP in subcarrier $k$, is
\begin{IEEEeqnarray}{rCl}
\label{eq_uplinksignal}
\hspace*{-20pt} \y_{\uplink} [k] = \sqrt{\frac{\rho}{K}} \PP_{\an}^T \HH^T[k] \g^{*} \x[k] + \z_{\uplink}[k] \quad \in \ \C^{N_{\rf} \times T},
\end{IEEEeqnarray}
where $\rho$ is the transmitted power, $\g \in \C^{M_{\ue} \times 1}$ is the column vector of beamforming coefficients (steering vector) at the STA, $\HH^T[k] \in \C^{M_{\ap} \times M_{\ue}}$ is the uplink MIMO channel matrix between the STA and AP arrays, $\z_{\uplink}[k]$ is a complex white Gaussian noise matrix whose elements have zero-mean and variance $\sigma_z^2/N_{\rf}$, and $\PP_{\an} \in \C^{M_{\ap} \times N_{\rf}}$ is the analog beamforming matrix defined as
\begin{IEEEeqnarray}{rCl}
\label{eq_analogprecoder}
\PP_{\an} = \left[ \p_{\an,1} \ \cdots \ \p_{\an, N_{\rf}}\right],
\end{IEEEeqnarray}
where the column vector $\p_{\an,n}$ represents the $n$-th RF chain beamforming coefficients. The time-domain sequence received at RF chain $n$ and subcarrier $k$ is
\begin{IEEEeqnarray}{rCl}
\label{eq_uplinksignalRF}
\y_{\uplink,n} [k] & = & v_n[k] \x[k] + \z_{\uplink,n}[k],
\end{IEEEeqnarray}
where
\begin{IEEEeqnarray}{rCl}
\label{eq_uplinkcoeff}
v_n[k] = \sqrt{\frac{\rho}{K}} \p_{\an,n}^T \HH^T[k] \g^{*}
\end{IEEEeqnarray}
is the complex channel coefficient after beamforming. We assume the power constraints $\left\Vert \p_{\an,n} \right\Vert_2^2 = N_{\rf}^{-1} \ \forall n$, $\left\Vert \PP_{\an} \right\Vert_F^2=1$, and $\left\Vert \g \right\Vert_2^2 = 1$, which take into account the power splitters/combiners in the system. This model, depicted in Fig. \ref{fig_modeltraining}, assumes that the transmitted power is uniformly distributed across subcarriers. Note that the analog beamformers $\PP_{\an}$ and $\g$, which represent the phase shifters configuration, are independent of frequency (we elaborate on this property in Section \ref{sec_phaseshifters}).

\subsection{Single-User mmWave Downlink - Beam Selection}
\label{sec_systemmodeltraininguplink}
For the downlink, the AP sends a training signal $\x [k]$ with equal power through all the RF chains and subcarriers. Using the hardware described above, and assuming a reciprocal channel, the received signal at the STA for subcarrier $k$ is
\begin{IEEEeqnarray}{rCl}
\label{eq_downlinksignal}
\y_{\dl}[k] & = & w[k] \x[k] + \z_{\dl}[k] \quad \in \ \C^{1 \times T},
\end{IEEEeqnarray}
where $\z_{\dl}[k] \thicksim \cgaussian \left(\ze,\sigma_z^2 \I_T \right)$ is complex white Gaussian noise, and the complex downlink channel coefficient is 
\begin{IEEEeqnarray}{rCl}
\label{eq_downlinkcoeff}
w[k] & = & \sqrt{\frac{\rho}{K N_{\rf}}} \g^H \HH[k] \PP_{\an} \, \one, \nonumber \\
& = & \sqrt{\frac{\rho}{K N_{\rf}}} \g^H \HH[k] \left( \sum_{n=1}^{N_{\rf}} \p_{\an , n} \right).
\end{IEEEeqnarray}
where $\one$ is a $N_{\rf} \times 1$ vector whose entries are all 1. For convenience, we use the same total transmit power $\rho$ in the uplink and downlink, and we define the signal-to-noise ratio in the system as
\begin{IEEEeqnarray}{rCl}
\label{eq_snr}
\snr = \frac{\rho}{\sigma_z^2}.
\end{IEEEeqnarray}
Fig. \ref{fig_modeltraining} shows the downlink model and its relationship with the uplink. In the special case where all RF chains use the same beamformer $\p$, the downlink channel coefficient reduces to
\begin{IEEEeqnarray}{rCl}
\label{eq_downlinkcoeff2}
w[k] = \sqrt{\frac{N_{\rf}\rho}{K}} \g^H \HH[k] \p.
\end{IEEEeqnarray}
If we set $\p_{\an,n} = \p$ in (\ref{eq_uplinkcoeff}), the relationship $v[k] = N_{\rf}^{-\frac{1}{2}} w[k]$ between uplink and downlink channel coefficients holds due to channel reciprocity. The factor $N_{\rf}^{-\frac{1}{2}}$ accounts for the power loss caused by a single RF chain receiving with a given beamformer in the uplink, whereas all RF chains transmit the same training signal with the same beamformer in the downlink.

\subsection{Multiuser Downlink - Hybrid Beamforming}
When the AP sends independent signals to $U$ different users, the downlink is represented by a broadcast channel where the signal at user $u$ is
\begin{IEEEeqnarray}{rCl}
\label{eq_broadcastsignal}
y_u [k] & = & \h_{\eq,u}[k] \PP_{\di}[k] \s[k] + \z_{\dl,u}[k],
\end{IEEEeqnarray}
where $\PP_{\di}[k] \in \C^{ N_{\rf} \times U }$ is the digital precoder that maps the $U$ independent transmitted signals to the available RF chains, $\s[k] = \left[ s_1 [k], \ldots , s_U [k] \right]^T \in \C^{U \times 1}$ is the vector of complex symbols transmitted to the users, and $\h_{\eq,u}[k]$ is the equivalent MISO channel after analog beamforming for user $u$ defined as
\begin{IEEEeqnarray}{rCl}
\label{eq_equivalentchannel}
\h_{\eq,u}[k] & = & \g_u^H \HH_u[k] \PP_{\an} \quad \in \ \C^{1 \times N_{\rf}},
\end{IEEEeqnarray}
where $\g_u[k]$ and $\HH_u[k]$ are the beamformer and the channel matrix for user $u$, respectively, $\PP_{\an}$ is defined as in (\ref{eq_analogprecoder}). $\h_{\eq,u}[k]$ represents the channel coefficients between each RF chain and user $u$. The transmitted signals satisfy the total power constraint $\sum_u \rho_u \leq \rho$, where $\rho_u = \E \left[ \left| s_u[k] \right|^2 \right]$. In this model, the AP applies both analog and digital beamformers to the signal, so the total transmitter processing is represented by a hybrid precoding matrix $\PP_{\an} \PP_{\di}[k] \in \C^{M_{\ap} \times U}$. Thus, the maximum number of users the AP can support is limited by the number of RF chains ($U \leq N_{\rf}$). In addition, we use the power constraint $\left\Vert \PP_{\an} \PP_{\di}[k] \right\Vert_F = 1$, $\forall k$, which allows a fair comparison with other techniques. In Section \ref{sec_problem}, we formulate the problem of selecting analog beamformers $\PP_{\an}$ and $\left\{ \g_u[k] \right\}$ that maximize the received SNR at every user and a digital beamformer $\PP_{\di}[k]$ that eliminates the remaining inter-user interference.

\subsection{Channel Model}
The hybrid beamforming algorithm presented in Section \ref{sec_algorithm} does not require knowledge of the channel matrix $\HH[k]$. {However, it is specifically designed for uniform (linear or planar) antenna arrays. Thus, in this section we describe a simple channel model that is only instrumental to characterize the algorithm's performance through simulations. We stress that our algorithm is sufficiently general to operate in any channel that uses uniform arrays and without explicit channel information. Assuming $L$ propagation paths between the AP and the STA} (with angles of arrival and departure for path $\ell$ given by $\theta_{\ue,\ell}$ and $\theta_{\ap,\ell}$, respectively), the downlink channel matrix has the form
\begin{IEEEeqnarray}{rCl}
\label{eq_channelmodel}
\HH[k] & = & \left( \I_{M_{\ue}} + \SSS_{\ue}[k] \right) \nonumber \\
&& \times \left( \sum_{\ell=1}^L \alpha_\ell \aaa_{\ue} \left( k,\theta_{\ue,\ell} \right) \aaa_{\ap}^H \left( k,\theta_{\ap,\ell} \right) \right) \nonumber \\
&& \times \left( \I_{M_{\ap}} + \SSS_{\ap}[k] \right) \quad \in \ \C^{M_{\ue} \times M_{\ap}},
\end{IEEEeqnarray}
where $\SSS_{\ap}[k]$ and $\SSS_{\ue}[k]$ are the frequency-dependent S-parameter matrices of the antenna arrays at the AP and the STA, respectively, $\aaa_{\ap} \left( k,\theta_{\ap,\ell} \right) \in \C^{M_{\ap} \times 1}$ is the array response vector at the AP in the angle of departure $\theta_{\ap,\ell}$ at subcarrier $k$, $\aaa_{\ue} \left( k,\theta_{\ue,\ell} \right)\in \C^{M_{\ue} \times 1}$ is the array response vector at the user terminal in the angle of arrival $\theta_{\ue,\ell}$ at subcarrier $k$, and $\alpha_\ell \in \C$ is the coefficient of path $\ell$. This multipath model is adapted from \cite{sayeed2002} to consider antenna coupling effects by using the array's S-parameters as in \cite[Ch. 2]{mailloux2005}. {As an example for simulation purposes, we define the coupling between antennas $m$ and $m'$ in a ULA (assuming perfect antenna impedance matching) as}
\begin{IEEEeqnarray}{rCl}
\left[ \SSS [k] \right]_{m,m'} =
\begin{cases} 
0		&	\text{if } m = m', \\
c \frac{\exp \left( -j 2 \pi d \frac{f_k}{f_0} \left| m - m' \right| \right) } {\left| m - m' \right|}		&	\text{if } m \neq m',
\end{cases} \qquad
\end{IEEEeqnarray}
where $c$ is a scalar that determines the coupling amplitude, which is typically below $-15$ dB for adjacent antennas \cite{rohani2018}. The idea behind this model is to generate coupling coefficients whose power decays with the squared of the distance separating antenna elements. This is a conservative model for coupling, especially for adjacent elements that observe a sum of power components of higher order. However, the model can be easily adjusted using the constant $c$ to match measurements of implemented antenna arrays. {We note that this antenna coupling model is only used for testing in Section} \ref{sec_results} {and can be replaced with measurement data, if available.} The array response vectors depend exclusively on the array geometry. The model given by (\ref{eq_channelmodel}) assumes ULAs but it can also be extended to planar antenna arrays as pointed out in \cite{song2017}.

\subsection{Antenna Array Response Vector and Radiation Pattern}
One of the features of OFDM mmWave systems operating with the hardware configuration described above is that the array response vectors are frequency-dependent. {This is caused by changes in electrical lengths at different frequencies}. As a consequence of this feature, a fixed phase shifter configuration has different maximum radiation directions for different subcarriers. This effect is commonly known as \emph{beam squint} and can severely impact the performance of mmWave communications \cite{cai2016,cai2017}{. }To analyze its impact on the array's radiation pattern, let $f_0$ denote a reference frequency in the band of interest, such that the array inter-element spacing is referenced to it. Let $f_k$ denote the frequency of subcarrier $k$. As an example, we present a ULA model for the AP, where $\theta$ represents the angle with respect to the array axis. The elements in the array response vector $\aaa_{\ap} (k,\theta)$ represent the phase of an incoming/outgoing plane wave received/generated by the array in the far field \cite{balanis2016}. For a ULA, the $m$-th element in $\aaa_{\ap}(k,\theta)$ is
\begin{IEEEeqnarray}{rCl}
a_{\ap,m}(k,\theta) \hspace*{-2pt} = \hspace*{-2pt} F(k,\theta) \exp \left( j 2 \pi d \frac{f_k}{f_0} \left[ m \hspace*{-2pt} - \hspace*{-2pt} \frac{M \hspace*{-2pt} + \hspace*{-2pt} 1}{2} \right] \cos \theta \right) \hspace*{-2pt}, \nonumber \\
\end{IEEEeqnarray}
where $m=1,\ldots,M_{\ap}$, $d$ is the antenna element spacing (normalized to the central frequency wavelength), and $F(k,\theta)$. As an example, and with performance evaluation purposes, we use antenna elements with radiation patterns 
\begin{IEEEeqnarray}{cCl}
\label{eq_radiationpattern}
F(k,\theta) = 
\begin{cases}
2\sin\theta						& \text{if } \theta \in [0,\pi], \\
10^{-2} 						& \text{otherwise}. \\
\end{cases}
\end{IEEEeqnarray}
This model approximate mmWave antennas found in literature, e.g. \cite{hamberger2017,korn2017,sahin2016}, which are typically mounted over reflecting chassis. The arrays are able to scan a half-space only, with a small power leakage to the back. If suitable, this model can be replaced with measurement data for a more accurate performance evaluation. The array's electric field pattern at subcarrier $k$ is then
\begin{IEEEeqnarray}{rCl}
\Psi_{\ap}(k,\theta) = \aaa_{\ap}^H (k,\theta) \left( \I + \SSS_{\ap} [k] \right) \p.
\end{IEEEeqnarray}
Analogous definitions apply to the array's radiation pattern at the STA $\Psi_{\ue}(k,\theta)$.
\begin{figure}[t]
\centering
\subfloat[]{\hspace*{24pt}\includegraphics[height=1.1in]{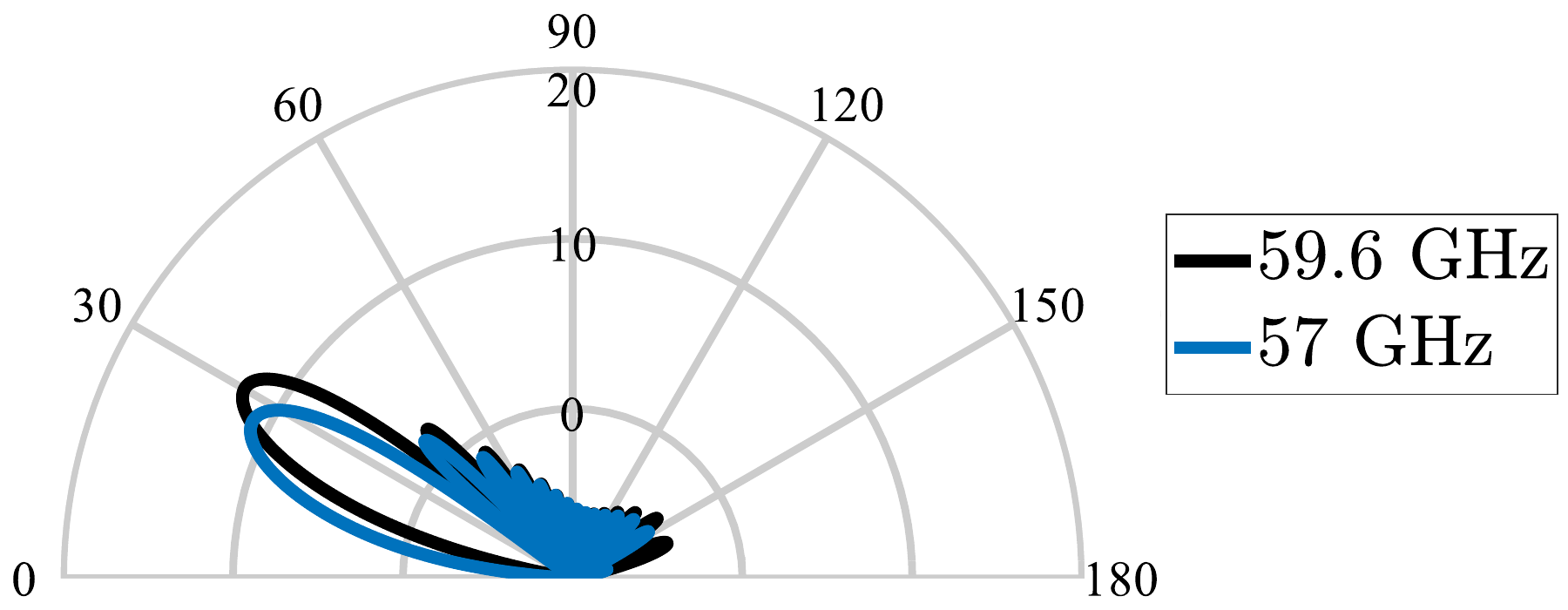}}\\
\subfloat[]{\hspace*{24pt}\includegraphics[height=1.1in]{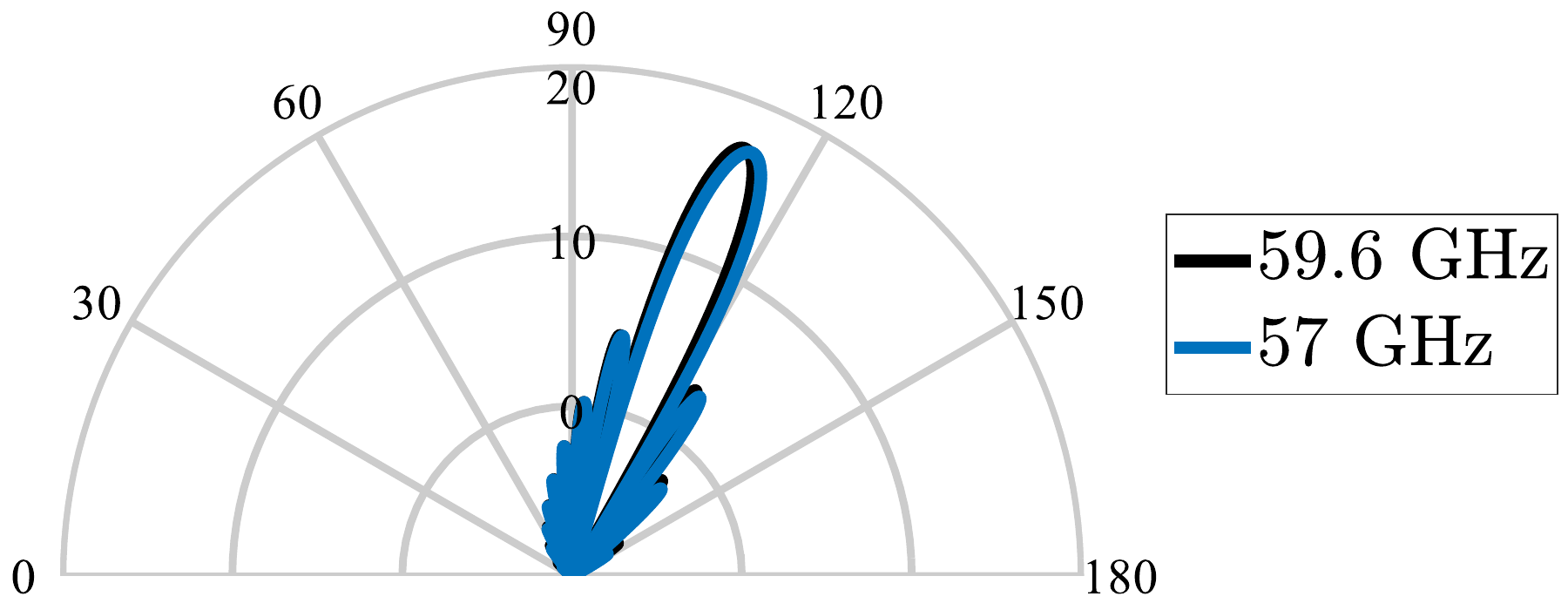}}
\caption{Power radiation pattern $\left| \Psi_{\ap} \left( k,\theta \right) \right|^2$ (in dBi) generated by a ULA with $M_{\ap} = 16$ and element spacing $d=0.5$ at $f_0 = 60$ GHz plotted for the highest and lowest subcarriers in Channel 1 for IEEE 802.11ad \cite{iso2014}. Fixed phase shifts for main beams pointing to (a) $\theta \approx 30^{\circ}$ and (b) $\theta \approx 110^{\circ}$ at $f_0$. \emph{Beam squint} is more pronounced for angles closer to the array's axis.}
\label{fig_beamsf}
\end{figure}
%When there is no antenna coupling and the antennas have perfect impedance matching, $\SSS_{\ap}[k] = \ze$ and the array's pattern reduces to
%\begin{IEEEeqnarray}{rCl}
%\Psi_{\ap}(k,\theta) = F (k,\theta,\phi) \ \aaa_{\ap}^H (k,\theta) \p. \nonumber
%\end{IEEEeqnarray}
% The phase is referenced to the array's center and is dependent on the subcarrier frequency $f_k$. Considering arbitrary antenna elements with radiation pattern $F(k,\theta)$, antenna coupling, and a beamformer vector $\p$ representing the phase shifter configuration at the antenna elements, t

\subsection{RF Phase Shifters}
\label{sec_phaseshifters}
Phase shifters commonly have a frequency-independent group delay $\tau$, which translates into a quasi-linear phase change with respect to frequency . A typical group delay value for mmWave phase shifters is around 10ps$-$100ps \cite{adabi2010,koh2008}. Thus, we can model the frequency response of a phase shifter as $b (k) = e^{-j 2 \pi \tau (f_k-f_0)} e^{ j \beta_0 }$, where $\beta_0$ is the phase shift at the reference frequency $f_0$. However, the phase difference between phase shifters connected to different antenna elements is independent of $e^{-j 2 \pi \tau (f_k-f_0)}$. Thus, we ignore this term when designing the beamforming vectors so that they are independent of frequency. Fig. \ref{fig_beamsf} shows an example of two radiation patterns at different frequencies in the 60 GHz band obtained with the models described in this Section and {using the parameters described in the figure caption}.

%% file: Problem.tex
\section{Problem Statement}
\label{sec_problem}

After defining the system model, {we now formally state our approach to beamforming for the wideband multiuser downlink in} (\ref{eq_broadcastsignal}). We propose to decouple the design of $\PP_{\an}$ and $\PP_{\di}[k]$. The general idea is to first find analog beamformer vectors that maximize the SNR for each STA and construct the analog matrix $\PP_{\an}$ using those vectors. Then, the AP can design $\PP_{\di}[k]$ with conventional digital beamforming techniques to eliminate the remaining inter-user interference. To construct $\PP_{\an}$, we aim to solve the following optimization problem for each user:
\begin{IEEEeqnarray}{rCl}
\label{eq_optimbeamselection1}
\left\{ \p_u^\star , \g_u^\star \right\} & = & \arg \max_{\substack{\p \in \Bset (M_{\ap}) \\ \g \in \Bset (M_{\ue}) }}  \sum_k  \left|  \g^H \HH_u[k] \p \right|^2, \nonumber \\
& = & \arg \max_{\substack{\p \in \Bset (M_{\ap}) \\ \g \in \Bset (M_{\ue}) }}  \left\Vert \vv_u \right\Vert^2 
\end{IEEEeqnarray}
where $\vv_u = \left[ v_u[1],\ldots,v_u[K] \right]^T$ is the channel coefficient vector, $v_u[k]$ is the channel coefficient as given in (\ref{eq_uplinkcoeff}) for user $u$, $ \Bset (M_{\ap})$ and $\Bset (M_{\ue})$ are discrete beamformer sets for the AP and STAs, respectively. These sets are known as \emph{codebooks} and their cardinality depends on the number of antennas and the quantized phase-shifter configurations. The optimization problem above can be understood as a sum-power maximization across all subcarriers, where the system selects the best beamformers available at the AP and the STA. Solving this optimization problem with exhaustive search requires perfect knowledge of the objective function for every combination of feasible values of $\p$ and $\g$. Since this is unattainable in practice due to noise, we relax the problem in (\ref{eq_optimbeamselection1}) by using a channel vector estimate $\vbh_u$ to get
\begin{IEEEeqnarray}{rCl}
\label{eq_optimbeamselection2}
\left\{ \p_u^\star , \g_u^\star \right\} & = & \arg \max_{\substack{\p \in \Bset (M_{\ap}) \\ \g \in \Bset (M_{\ue}) }}  \left\Vert \vbh_u \right\Vert^2.
\end{IEEEeqnarray}
Note that, unlike other methods, our algorithm does not attempt to obtain an estimate of the channel matrix $\HH_u[k]$ nor assumes perfect CSI, but instead it estimates the channel coefficients $\vv_u$ after beamforming, thus reducing the computational complexity. To solve this problem, we propose a heuristic algorithm {to decrease} the required number of training transmissions with respect to an exhaustive search over $\Bset (M_{\ap}) \times \Bset (M_{\ue})$. Our approach is based on hierarchical codebooks that we describe in Section \ref{sec_codebook}. Note that training transmissions provide maximum likelihood (ML) channel estimates for every user. However, using ML estimates, the solutions to (\ref{eq_optimbeamselection1}) and (\ref{eq_optimbeamselection2}) are equal only when the $\snr$ is asymptotically large. This fact, {combined with the algorithm's operation, causes a nonzero probability of obtaining different solutions to} (\ref{eq_optimbeamselection1}) and (\ref{eq_optimbeamselection2}) at finite $\snr$. If those solutions differ, we declare a beam selection error, and we use the \emph{beam selection error rate} (BSER) as an algorithm performance metric. Once (\ref{eq_optimbeamselection2}) is solved for every $u$, $\PP_{\an}$ is constructed using $\left\{ \p_u^\star \right\}$ as columns. Estimates of the equivalent MISO channel given by (\ref{eq_equivalentchannel}), and denoted by $\hbh_{\eq,u}[k]$ for user $u$, are obtained for every STA. Finally, the AP calculates the $\PP_{\di}[k]$ using block-diagonalization to eliminate the residual inter-user interference after analog beamforming. This algorithm is detailed in Section \ref{sec_algorithm}.
 
%Codebooks $\Pset_s$ and $\Pset$ have \emph{sector} beams and \emph{narrow} beams for $\PP_{\an}$, respectively, while Similarly, $\Gset_s$ and $\Gset$ are the sector and narrow beamformers for $\g$. These codebooks are designed to operate in the training modes given by (\ref{eq_uplinksignalRF}) and (\ref{eq_downlinksignal}). 

%% file: Codebook.tex
\section{Codebook Design}
\label{sec_codebook}

In this section, we describe a method to design beamforming codebooks for the analog beamformers $\PP_{\an}$ and $\left\{ \g_u \right\}$ based on the orthogonality of beamforming vectors. We leverage the hardware architectures described in Section \ref{sec_sytemmodel} to construct hierarchichal codebooks whose elements provide main beams of two widths. The first codebook at the AP, denoted as $\Pset_s$, scans wide \emph{sectors} of the angular domain in the uplink by using different (adjacent) narrow beams in each RF chain. At the STAs, $\Gset_s$ is the codebook with \emph{sector} beamformers obtained using the subarray configuration. The codebooks with \emph{narrow} (pencil) beams are denoted by $\Pset$ and $\Gset$ at the AP and the STAs, respectively. The algorithm in Section \ref{sec_algorithm} uses these codebooks to solve (\ref{eq_optimbeamselection2}).
\subsection{Orthogonal Beamformers for ULAs}
\begin{figure}[t]
\centering
\subfloat[]{\includegraphics[height=1.1in]{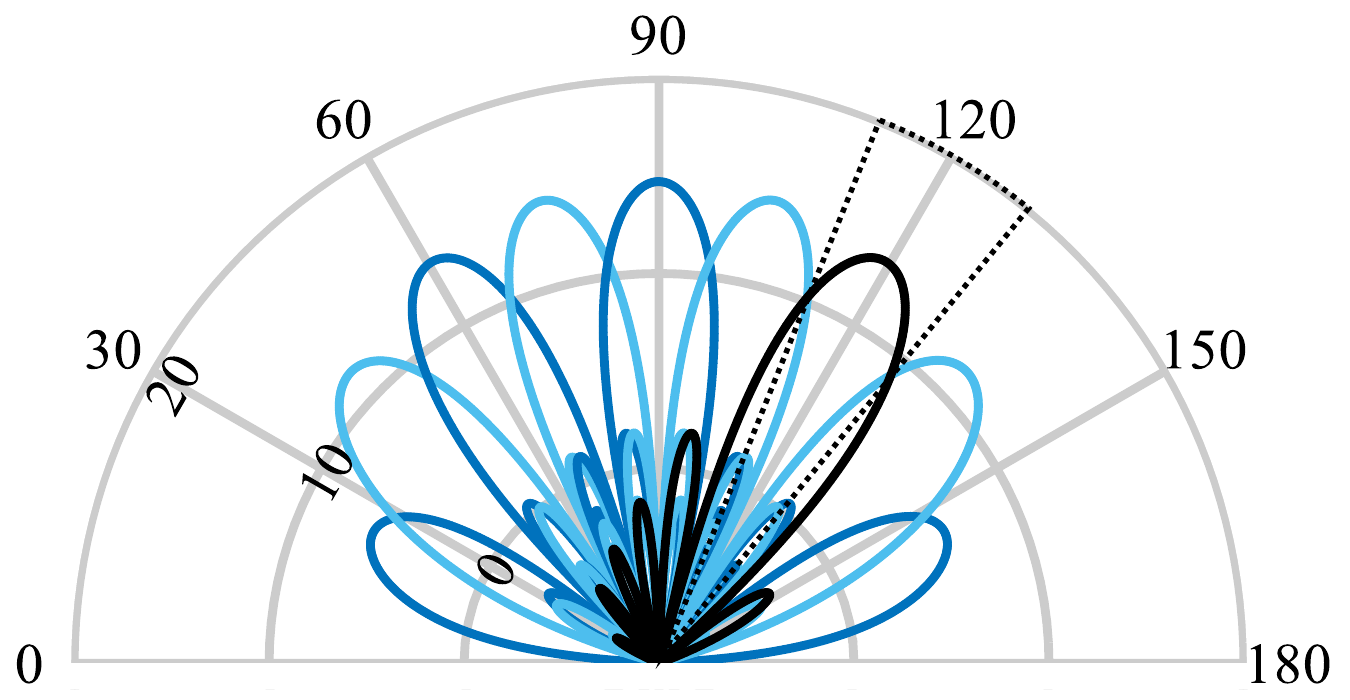}}\\
\subfloat[]{\includegraphics[height=1.1in]{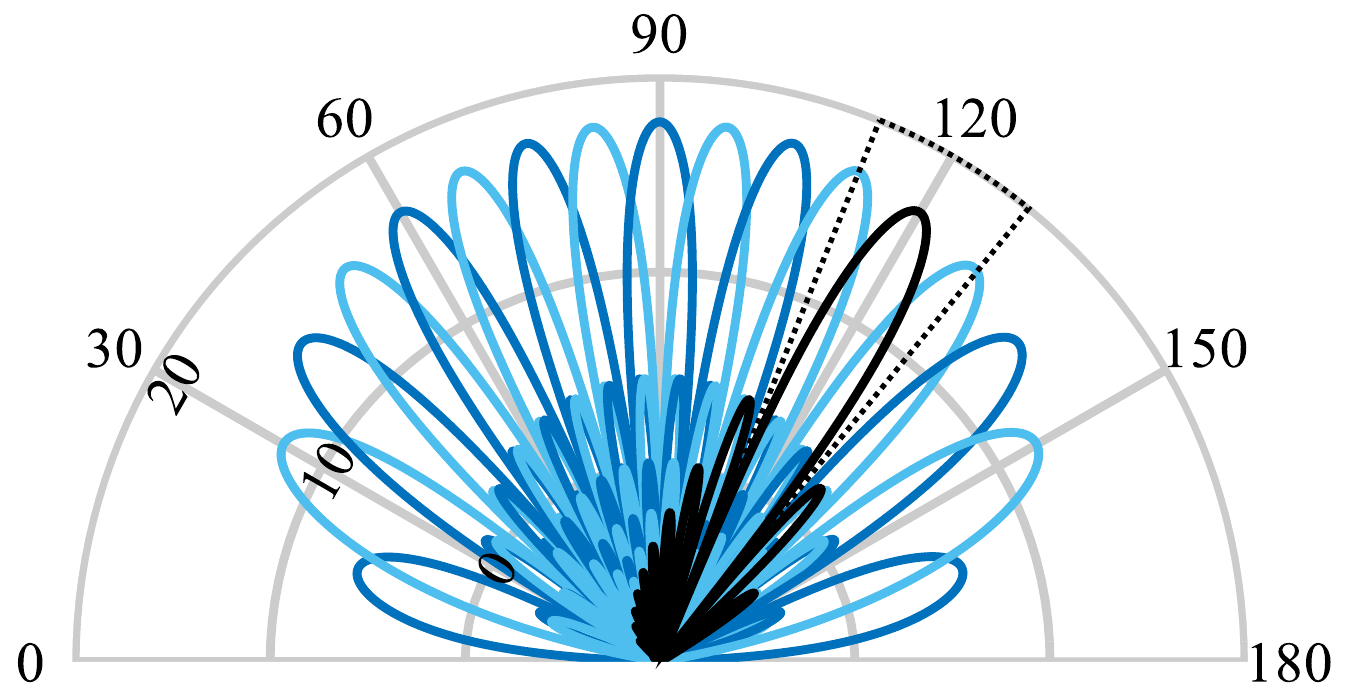}}
\caption{Power radiation patterns (in dBi) generated by orthogonal beamformers for for a uniform linear array with (a) 8 antennas ($\Bset(8)$) with $\bb_7(8)$ highlighted in black, and (b) 16 antennas ($\Bset(16)$) with $\bb_{13}(16)$ highlighted in black, both operating at the central frequency. Parameters: $d=0.5$, $F(k,\theta) = 2 \sin \theta$. The first beamformer is symmetric with one grating lobe. The sector spanned by $\bb_7(8)$ (dotted line) overlaps with beams generated by $\bb_{12}(16)$, $\bb_{13}(16)$, and $\bb_{14}(16)$.}
\label{fig_beams816}
\end{figure}
We construct codebooks using orthogonal beamforming vectors for uniform arrays. Orthogonal beamformers allow an efficient exploration of the channel subspaces \cite{alkhateeb2016}. {They also have interesting properties for hybrid and subarray configurations}. Orthogonal beamformers simplify the analog precoder design, facilitate the hierarchical structure design, and enable a frequency-independent alignment between \emph{narrow} beams and \emph{sector} beams. Thus, the constructed codebooks are not vulnerable to beam squint. With the goal of simplifying the codebook description, we assume that the AP and the STAs have 1D ULAs printed over a planar substrate with element spacing of half-wavelength at the frequency $f_0$, i.e. $d=0.5$. This configuration is common in mmWave systems\cite{abari2016}, and it allows us to describe the radiation pattern for $\theta \in [0,\pi)$ only. The orthogonality property also exists for uniform 2D antenna arrays, so the principles described here can be easily extended to other uniform array geometries. 

{Let $\Bset(M)$ denote a set of orthogonal beamforming vectors for ULAs with $M$ antennas defined as}
\begin{IEEEeqnarray}{rCl}
\label{eq_orthogonalset}
\Bset(M) = \left\{ \bb_m(M) \in \C^{M} \ | \ m=1,\ldots, M \right\}.
\end{IEEEeqnarray}
where
\begin{IEEEeqnarray}{rCl}
\hspace*{-5pt} \bb_m(M) = \frac{1}{\sqrt{M}} \left[ 1, e^{ j \beta_m(M) }, \ldots , e^{ j (M-1) \beta_m(M) } \right]. \quad
\end{IEEEeqnarray}
The entries in the beamforming vector represent the phase shift applied to each antenna element. The phase difference between elements is
\begin{IEEEeqnarray}{rCl}
\label{eq_orthogonalphases}
\beta_m(M) = \pi \left( 1 - 2 \frac{(m-1)}{M} \right).
\end{IEEEeqnarray}
The elements in $\Bset(M)$ are orthogonal, since they satisfy
\begin{IEEEeqnarray}{rCl}
\bb_m^H(M) \bb_{m'}(M) = \delta_{m,m'}, \ \forall M,
\end{IEEEeqnarray}
where $\delta_{m,m'}$ is the Kronecker delta function. Fig. \ref{fig_beams816} shows radiation patterns generated by two orthogonal beamformer sets, $\Bset(8)$ and $\Bset(16)$. Some important properties of the beamformers in $\Bset(M)$ and their radiation patterns are:
\begin{enumerate}
\item {Orthogonal beamformers explore the dominant subspaces of the channel matrix more efficiently} (see \cite[Sec. VI.A.]{alkhateeb2016} for a discussion of this property). {Moreover, their orthogonality is frequency-independent, so it is preserved even under beam squint.}
\item {Another convenient property of the beamformers defined in }(\ref{eq_orthogonalset}) and (\ref{eq_orthogonalphases}) {is that the main beams generated by vectors in $\Bset(M_{\sub})$ (sectors) and the main beams generated by vectors in $\Bset(M_{\ue})$ (narrow beams) overlap in the angular domain.} More precisely, if $M_{\ue}$ and $M_{\sub}$ are two integers such that $M_{\ue}>M_{\sub}$ and $\frac{M_{\ue}}{M_{\sub}}$ is even, then there are precisely $\frac{M_{\ue}}{M_{\sub}}+1$ beams in $\Bset(M_{\ue})$ that overlap a wider beam in $\Bset(M_{\sub})$. This is observed in the sectors marked in Figs. \ref{fig_beams816} and \ref{fig_beamssub}, where $M_{\ue}=16$, $M_{\sub}=8$, and thus 3 narrow beams overlap with each sector beam. Furthermore, this property holds even when changing the operating frequency. We use this fact to construct hierarchical beamforming codebooks that are not affected by beam squint.
\item $\Vert \bb_m(M) \Vert_2 = 1$, $\forall m,M$ to ensure power normalization.
\item The set always contains a broadside beam (with direction of maximum radiation at $90^{\circ}$ with respect to the array axis) when $m=\tfrac{M}{2}+1$ independently of 
frequency and element spacing.
\item Their directions of maximum radiation are not uniformly distributed in the interval $[0,\pi)$. Beams which are closer to the array's axis are wider than beams close to the array's broadside.
\item The number of orthogonal beams is the same as the number of antennas. However, they do not provide the same gain or number of sidelobes since these features depend on the element pattern and spacing.
%\item If the element spacing at the operating frequency is larger than half-wavelength, {grating lobes} exist in the radiation pattern for one or more beams in the set. However, orthogonality ensures that the maximum radiation directions of a given beamformer (including grating lobes) coincide with nulls in the radiation patterns of orthogonal beamformers.
\end{enumerate}
\color{black}
We use the characteristics above to construct hierarchical codebooks for the AP and the STAs.
%\begin{IEEEeqnarray}{rCl}
%\bb_m(M) = \frac{1}{\sqrt{M}} 
%\begin{bmatrix}
%1	\\
%e^{ j \beta_m(M) }	\\
%\vdots					\\
%e^{ j (M-1) \beta_m(M) }	\\
%\end{bmatrix}. \nonumber
%\end{IEEEeqnarray}

\subsection{Sector Codebook with Hybrid Beamforming}
\begin{figure}[t]
\centering
\subfloat[]{\includegraphics[height=1.1in]{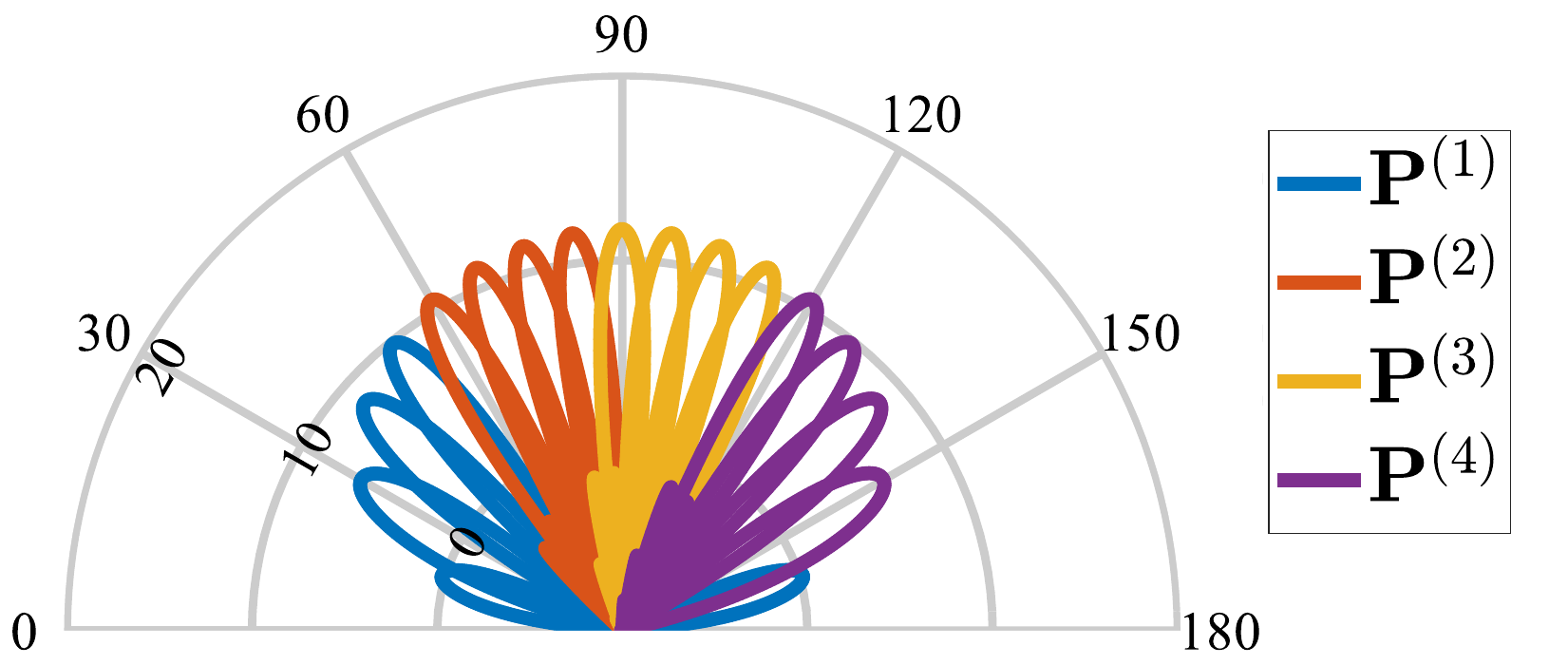}\label{fig_sectorbeamsrf}}\\
\subfloat[]{\includegraphics[height=1.1in]{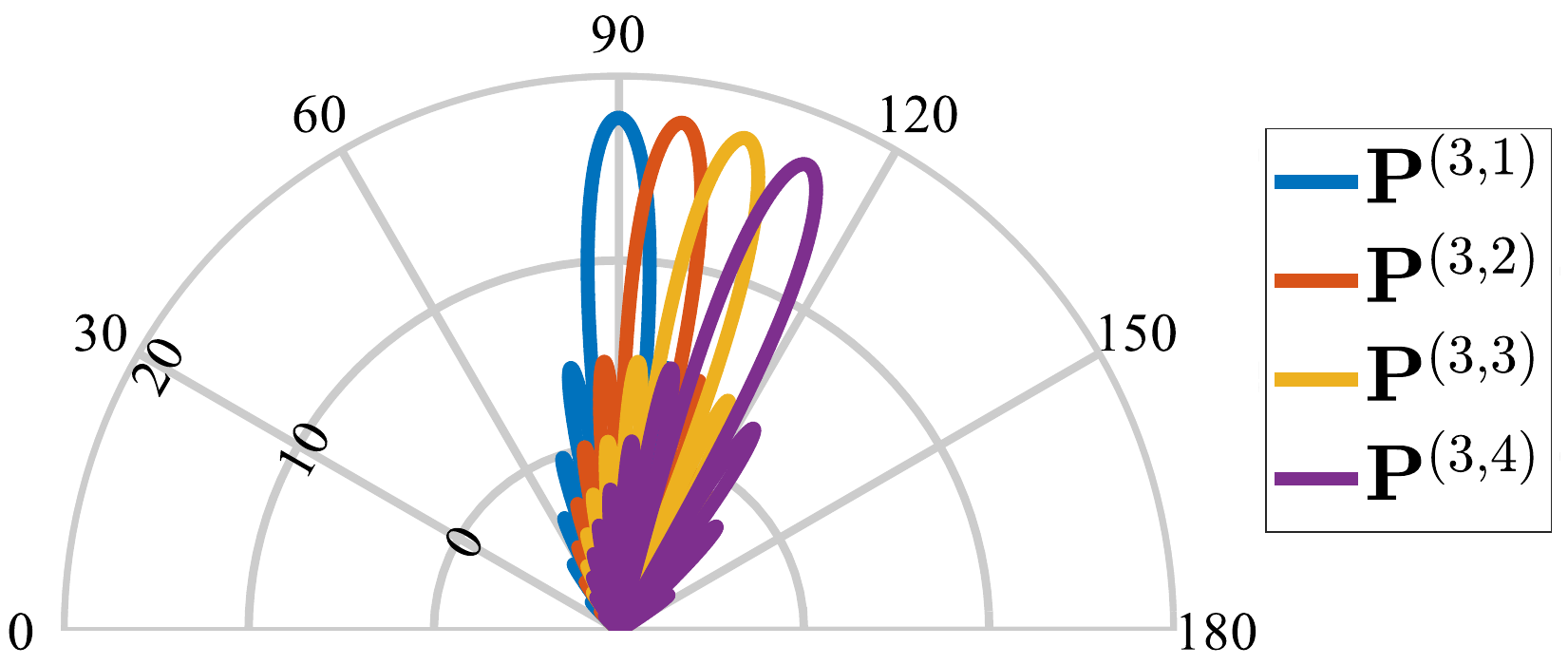}\label{fig_narrowbeamsrf}}
\caption{Power radiation patterns (in dBi) generated by (a) \emph{sector} beams $-$ RF chains in the uplink use adjacent narrow beam, and (b) \emph{narrow} beams $-$ all RF chains in the downlink use the same beamforming vector, thus increasing the array gain by an $N_{\rf}$ factor. In this example, $M_{\ap}=16$ antennas and $N_{\rf}=4$.}
\label{fig_beamsrf}
\end{figure}

In this section, we describe how to construct analog beamforming matrices that scan wide sectors in the angular domain by leveraging the hybrid architecture at the AP. These matrices form a codebook that contains possible configurations for $\PP_{\an}$ in the uplink given by (\ref{eq_uplinksignal}). We begin by defining the orthogonal $M_{\ap} \times N_{\rf}$ beamforming matrices as
\begin{IEEEeqnarray}{cCl}
\label{eq_orthogonalmatrix}
\B_m = \left[ \bb_{l_1(m)}(M_{\ap}),\bb_{l_2(m)}(M_{\ap}),\cdots,\bb_{l_{N_{\rf}}(m)}(M_{\ap})\right], \nonumber \\
l_n(m) = (m-1)N_{\rf} + n, \nonumber \\
m = 1,\ldots , M_{\ap} / N_{\rf}, \nonumber \\
n = 1, \ldots , N_{\rf}.
\end{IEEEeqnarray}
where the columns are $N_{\rf}$ elements in $\Bset(M_{\ap})$ with adjacent beams. The codebook of \emph{sector} beamforming matrices is then constructed as
\begin{IEEEeqnarray}{rCl}
\label{eq_codebookP1}
\Pset_s & = & \left\{ \PP^{(m)}=N_{\rf}^{-\frac{1}{2}}\B_m \ | \ m = 1, \ldots , M_{\ap} \right\},
\end{IEEEeqnarray}
where the factor $N_{\rf}^{-\frac{1}{2}}$ accounts for the $N_{\rf}$-port power splitter at each antenna. From the point of view of hardware configuration, each RF chain in the AP uses a different beamformer, such that different angles are seen at the uplink outputs. Fig. \ref{fig_sectorbeamsrf} shows an example of the radiation patterns for the sector codebook with $M_{\ap}=16$ antennas and $N_{\rf}=4$ chains.

\subsection{Narrow Beam Codebook with Hybrid Beamforming}
The codebook $\Pset$ has configurations for $\PP_{\an}$ that provide \emph{narrow} main beams for the downlink in (\ref{eq_downlinksignal}). Beamformers are constructed by configuring the same beamforming vector in all RF chains, which {directs} the radiated power in a narrow angular region. Codeword $n$ within sector $m$ is
\begin{IEEEeqnarray}{rCl}
\label{eq_codebookP2}
\PP^{(m,n)} = \frac{1}{\sqrt{N_{\rf}}} \left[ \bb_{l_n(m)}(M_{\ap})	\cdots \bb_{l_n(m)}(M_{\ap}) \right], \quad
\end{IEEEeqnarray}
where $l_n(m)$ is as defined in (\ref{eq_orthogonalmatrix}). This implies that the same beamforming vector $\bb_{l_n(m)}(M_{\ap})$ is assigned to all RF chains. The second-level codebook is then
\begin{IEEEeqnarray}{rCl}
\Pset & = & \left\{ \PP^{(m,n)} \ | \ (m,n) \in [ 1, ... \, , M_{\ap} ] \times [ 1, ... \, , N_{\rf} ] \right\}. \qquad
\end{IEEEeqnarray}
Fig. \ref{fig_narrowbeamsrf} shows a set of narrow beamformers $\left\{ \PP^{(3,n)} \right\}$, whose main beams point in directions within the sector beamformer $\PP^{(3)}$. Note the gain improvement by a $N_{\rf}$ factor with respect to the sector beamformers.
\begin{figure}[t]
\centering
\subfloat[]{\includegraphics[width=2.6in]{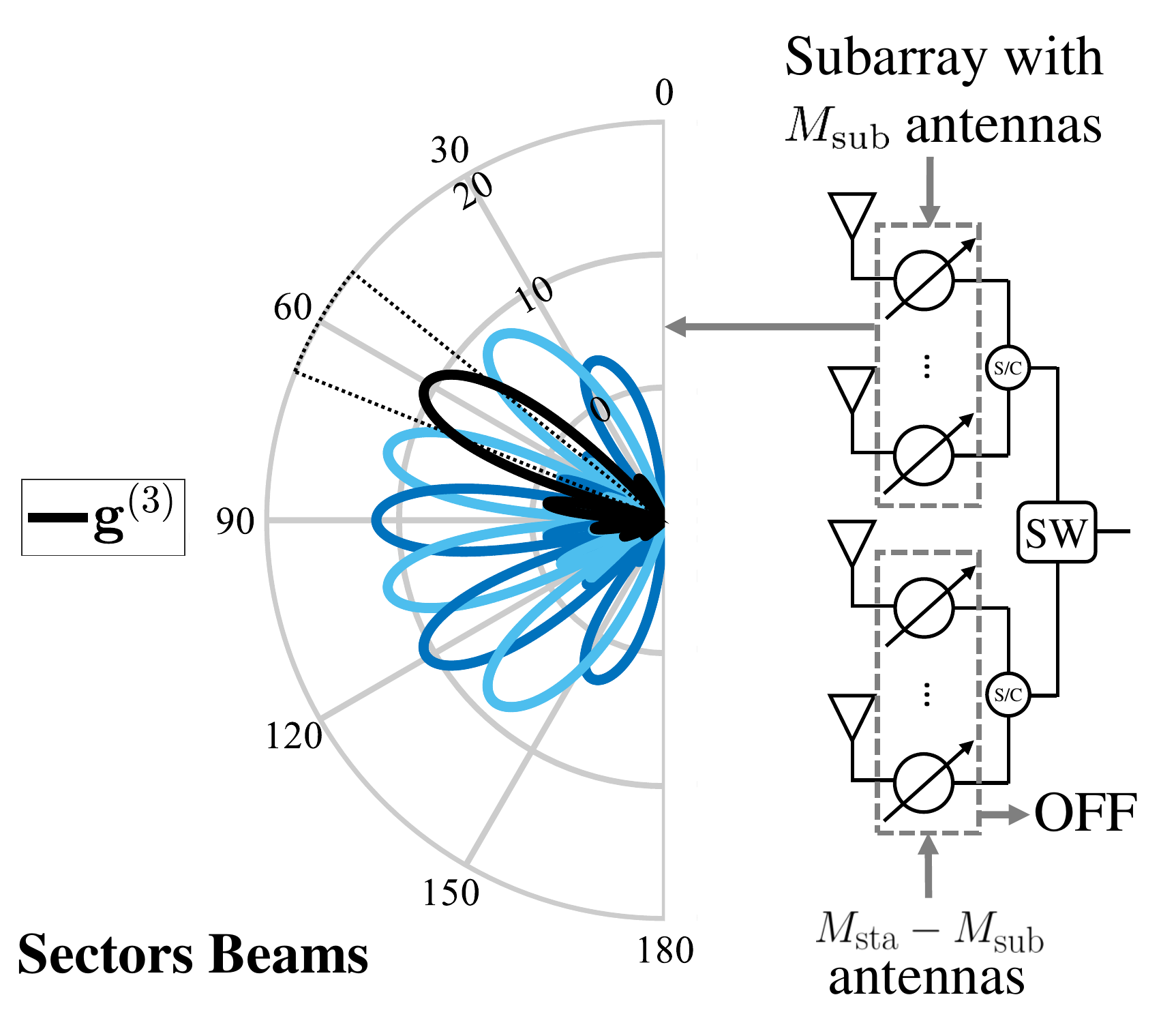}}\\
\subfloat[]{\includegraphics[width=2.6in]{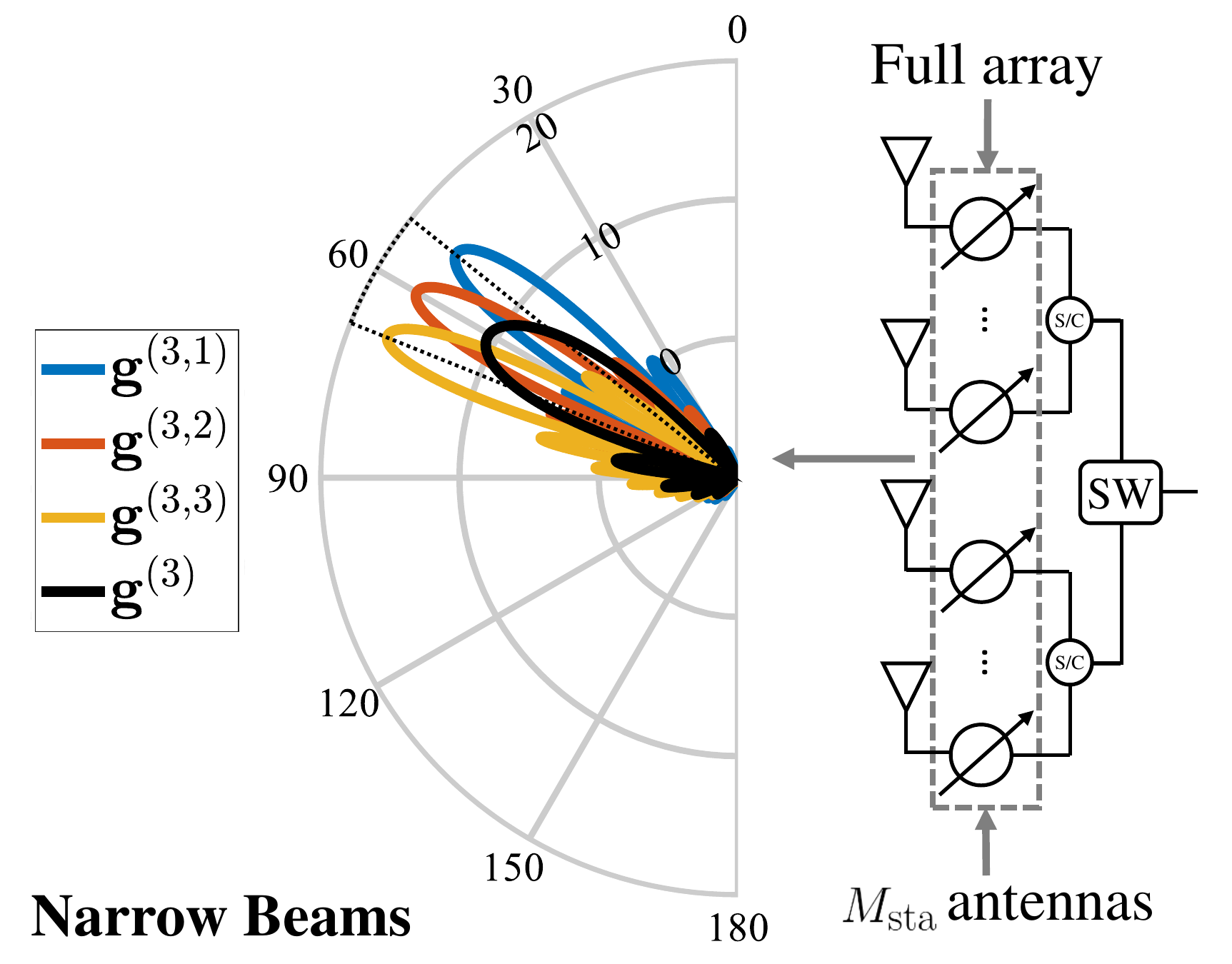}}
\caption{(a) Sector beams constructed from orthogonal beamformers when only the subarray is active ($\g^{(3)}$ is highlighted). (b) Narrow beams (full array configuration) that overlap with the sector $\g^{(3)}$. In this example, $M_{\ue}=16$ and $M_{\sub}=8$.}
\label{fig_beamssub}
\end{figure}
% 58/42 width ratio

\subsection{Sector Codebook with a Subarray}
We denote the sector codebook at the STAs as $\Gset_s$, which is used to select $\g$ to operate both in the uplink and downlink ((\ref{eq_uplinksignal}) and (\ref{eq_downlinksignal}), respectively). Sector beamformers are available only when the  $M_{\sub}$-antenna subarray is active at the STA. We use the orthogonal beamformer set $\Bset \left( M_{\sub} \right)$ to construct the \emph{sector} codebook as
\begin{IEEEeqnarray}{rCl}
\label{eq_codebookG1}
\Gset_s = \left\{ \g^{(m)} = \sqrt{\frac{M_{\sub}}{M_{\ue}}}
\begin{bmatrix}
\bb_m \left( M_{\sub} \right) \\
\ze\\
\end{bmatrix}, \ m = 1 , \ldots ,M_{\sub} \right\}, \nonumber \\
\end{IEEEeqnarray}
where the factor $\sqrt{\frac{M_{\sub}}{M_{\ue}}}$ accounts for the power loss in the switching network that activates the subarray.

\subsection{Narrow Beam Codebook with a Subarray}
The narrow beam codebook $\Gset$ is also used to select $\g$ at the STA, and is constructed using narrow orthogonal beams from $\Bset(M_{\ue})$. This means that the STA uses the whole antenna array. We define the codebook as 
\begin{IEEEeqnarray}{rCl}
\Gset = \left\{ \g^{(m,n)} \ | \ (m,n) \in [ 1, ... \, , M_{\ap} ] \times \left[ 1, ... \, , \tfrac{M_{\ue}}{M_{\sub}}+1 \right] \right\}, \nonumber
\end{IEEEeqnarray}
where the codewords are selected as
\begin{IEEEeqnarray}{cCl}
\g^{(m,n)} = \bb_{l(m,n)} (M_{\ue}), \\
l(m,n) = \left[ \tfrac{M_{\ue}}{M_{\sub}}(m-1) - \tfrac{M_{\ue}}{2M_{\sub}} + n \right]_{\bmod{M_{\ue}}}, \nonumber 
\end{IEEEeqnarray}
and we define the function
\begin{IEEEeqnarray}{cCl}
\left[ n \right]_{\bmod{M}} = 
\begin{cases}
M,				& \text{if } n\bmod M = 0, \\
n \bmod M 		& \text{if } n\bmod M \neq 0. \\
\end{cases}\qquad
\end{IEEEeqnarray}
This definition guarantees that, for a fixed $m$, beamformers in the set $\left\{ \g^{(m,n)} \right\}_{n=1}^{\frac{M_{\ue}}{M_{\sub}}+1}$ point their main beams in directions overlapping the sector beam generated by $\g^{(m)}$. The function $\left[ \cdot \right]_{\bmod{M}}$ is required to include the  adjacency between the main beams generated by $\bb_{1}(M_{\ue})$ and $\bb_{M_{\ue}}(M_{\ue})$. Fig. \ref{fig_beamssub} shows the construction of codebooks $\Gset_s$ and $\Gset$ with $M_{\ue}=16$ antennas and $M_{\sub} = 8$ antennas. Note that the 3 beamformers in $\left\{ \g^{(3,n)} \right\}$ overlap with the sector from $\g^{(3)}$.

%% file: Algorithm.tex
\section{mmWave OFDM Beamforming Algorithm}
\label{sec_algorithm}

In this section, we present an algorithm for the independent construction of $\PP_{\an}$ and $\PP_{\di}[k]$ in the multiuser downlink given by (\ref{eq_broadcastsignal}). The algorithm also finds beamformers $\left\{ \g_u \right\}$ for every STA and estimates their equivalent channel coefficients.

%%%%%%%%%%%%%%%%%%%%%%%%%%%%%%%%%%%%%%%%%%%%%%%%%%%%%%%%%%%%%%%%%%%%%%%%%%%%%%%%%%%%%%%%%%%%%%%%%
%%%%%%%%%%%%%%%%%%%%%%%%%%%%%%%%%%%%%%%%%%%%%%%%%%%%%%%%%%%%%%%%%%%%%%%%%%%%%%%%%%%%%%%%%%%%%%%%%
%%%%%%%%%%%%%%%%%%%%%%%%%%%%%%%%%%%%%%%%%%%%%%%%%%%%%%%%%%%%%%%%%%%%%%%%%%%%%%%%%%%%%%%%%%%%%%%%%

\subsection{General Description of the Algorithm}

\begin{figure*}[t]
\centering
\includegraphics[width=1.6\columnwidth]{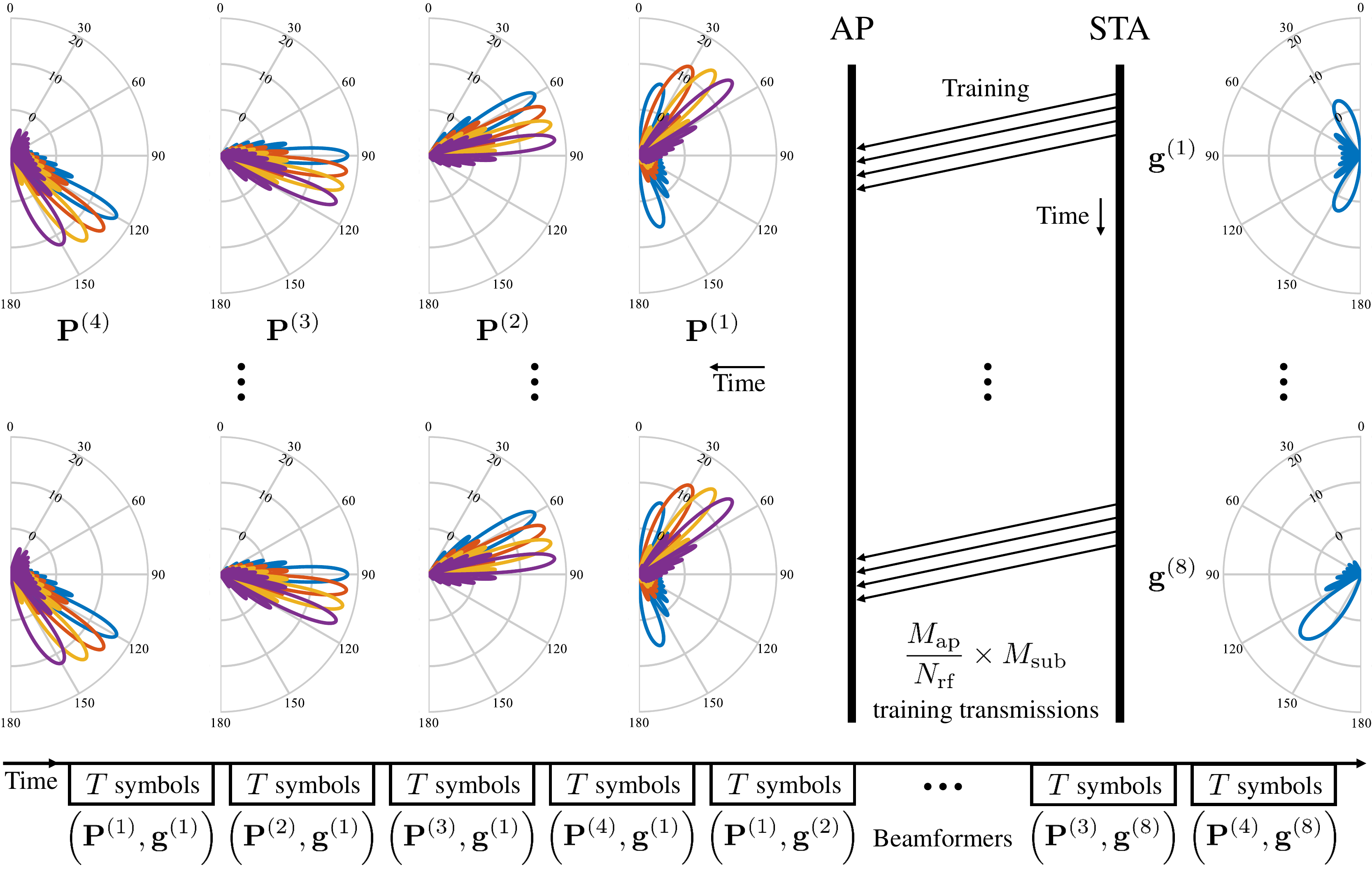}
\caption{Beam selection procedure - First stage (uplink). The STA transmits the training signal using every sector beamformer in $\Gset_s$. For each STA beamformer, the AP tries every sector beamformer in $\Pset_s$. In this example, $M_{\ap} = M_{\ue} = 16$ antennas, $M_{\sub}=8$ antennas, and $N_{\rf}=4$ chains.}
\label{fig_stage1}
\end{figure*}

Our algorithm for hybrid beamforming is split into two procedures: \emph{i}) the analog beamforming design procedure (or \emph{beam selection}) to solve (\ref{eq_optimbeamselection2}), and \emph{ii}) the digital beamformer design based on block-diagonalization (BD). \emph{Beam selection} is applied to every STA independently and consists of 3 training transmission alternating between uplink (first) and downlink (second and third). This is equivalent to an alternating optimization of the problem in (\ref{eq_optimbeamselection2}) where, starting with a fixed $\g$ (Stage 1), the algorithm solves for $\p$ using an exhaustive search over $\Pset_s$. Given the hierarchical codebook structure and the fact that $\Pset_s$ contains all the beamformers in $\Bset(M_{\ap})$ distributed over different RF chains, a search over $\Pset_s$ is equivalent to a search over $\Bset(M_{\ap})$. Once an optimum configuration for $\PP_{\an}$ is found, the AP begins training transmissions in the downlink while the STA performs an exhaustive search over $\Gset_s$ (Stage 2) and then a limited search over $\Gset$ (Stage 3). The limited search uses the superposition property of hierarchical orthogonal beamformers $\Gset_s$ and $\Gset$. When the training transmissions are finished and beamformers for all STAs are selected, we describe how to construct $\PP_{\an}$ using the beamformers to each STA as columns. The digital beamforming design requires channel estimates $\left\{ \hbh_{\eq,u} \right\}$ obtained with training transmissions from each STA, where the AP uses $\PP_{\an}$ as constructed in the previous stage. Once these estimates are available, the AP applies BD to calculate $\PP_{\di}[k]$ as described in \cite{spencer2004}. The details of each stage are presented next.

%%%%%%%%%%%%%%%%%%%%%%%%%%%%%%%%%%%%%%%%%%%%%%%%%%%%%%%%%%%%%%%%%%%%%%%%%%%%%%%%%%%%%%%%%%%%%%%%%
%%%%%%%%%%%%%%%%%%%%%%%%%%%%%%%%%%%%%%%%%%%%%%%%%%%%%%%%%%%%%%%%%%%%%%%%%%%%%%%%%%%%%%%%%%%%%%%%%
%%%%%%%%%%%%%%%%%%%%%%%%%%%%%%%%%%%%%%%%%%%%%%%%%%%%%%%%%%%%%%%%%%%%%%%%%%%%%%%%%%%%%%%%%%%%%%%%%

\subsection{Beam Selection - Stage 1 (Uplink)}

{This stage is summarized in Algorithm} \ref{alg_beamselection1}{. Fig. }\ref{fig_stage1} {shows an implementation example.} This stage's goal is to find beamformer $\p^\star \in \Bset(M_{\ap})$ that solves (\ref{eq_optimbeamselection2}). The detailed steps are the following:

\begin{enumerate}

\item The STA uses every sector beamformer in $\Gset_s$ to transmit $\frac{M_{\ap}}{N_{\rf}}$ training sequences sequentially. For each sector beamformer at the STA, the AP receives the training signals by sequentially sweeping through its $\frac{M_{\ap}}{N_{\rf}}$ sector beams.

\item Let $\y_{\uplink,n}^{m,m'} [k]$ be the uplink received signal at RF chain $n$ when $\g = \g^{(m')}$ and $\PP_{\an} = \PP^{(m)}$ (see (\ref{eq_uplinksignalRF})). We denote the corresponding channel coefficient as $v_{n}^{m,m'}[k]$. Since $\y_{\uplink,n}^{m,m'} [k]$ is a $T$-dimensional Gaussian row vector, {we obtain} an ML channel coefficient estimation as \cite[Sec. 4.4]{levy2008} 
\begin{IEEEeqnarray}{rCl}
\label{eq_stage1channelestimate}
\vh_{n}^{m,m'}[k] =\frac{\y_{\uplink,n}^{m,m'}[k] \x^H[k]}{\Vert \x[k] \Vert_2^2}.
\end{IEEEeqnarray}
To reduce the computational complexity, this channel estimation can also be performed for a reduced subcarrier subset $\Kset \subseteq \{ 1,\ldots,K \}$. The subcarriers used for channel estimation are commonly known as \emph{pilot subcarriers} and they can be used to track changes in the channel while other subcarriers are used for data transmission. We denote the number of pilot subcarriers as $ K_{\tx} = |\Kset|$ and analyze the algorithm's performance with respect to $K_{\tx}$ in Section \ref{sec_results}. The vector of channel estimates is $\vbh_{n}^{m,m'} = \left[ \vh_{n}^{m,m'}[1], \ldots , \vh_{n}^{m,m'}[K] \right]^T$.

\begin{algorithm}[t]
\caption{Analog Beam Selection - Stage 1 (Uplink)}
\begin{algorithmic}[1]
\label{alg_beamselection1}
\renewcommand{\algorithmicrequire}{\textbf{Input:}}
\renewcommand{\algorithmicensure}{\textbf{Output:}}
\REQUIRE $\Pset_s$ known at the AP and $\Gset_s$ known at the STA. Training signal $\x[k]$.
\ENSURE  Beamformers $\PP^\star$, $\p^\star$ known at the AP.
\FOR {$m' = 1$ to $M_{\sub}$}
	\STATE STA: set $\g = \g^{(m')}$.
	\FOR {$m = 1$ to $\tfrac{M_{\ap}}{N_{\rf}}$}
		\STATE STA: Transmit $\x[k]$.
		\STATE AP: set $\PP_{\an} = \PP^{(m)}$ and receive the signal.
		\FOR {every $k \in \Kset$}
			\STATE AP: estimate $v^{m,m'}_{n}[k]$ using (\ref{eq_stage1channelestimate}).
		\ENDFOR
	\ENDFOR
\ENDFOR
\STATE AP: obtain $\PP^\star$ and $\p^\star$ using (\ref{eq_optimstage1}) and (\ref{eq_algorithmsolutionAP}).
\RETURN  $\PP^\star$ and $\p^\star$ known at the AP.
\end{algorithmic} 
\end{algorithm}

%\begin{IEEEeqnarray}{rCl}
%\label{eq_stage1channelcoeff}
%\hspace*{-50pt} v_{m,n}^{m'}[k] & = & \sqrt{\frac{\rho}{K N_{\rf}} } \left[ \B_{\an,m}[k] \right]_{:,n}^T \HH^T[k] \frac{ \vect \left( \B_{\sub,m'}^*[k] \right) }{\sqrt{N_{\sub}}}. 
%\end{IEEEeqnarray}

\begin{figure*}[t]
\centering
\includegraphics[width=1.6\columnwidth]{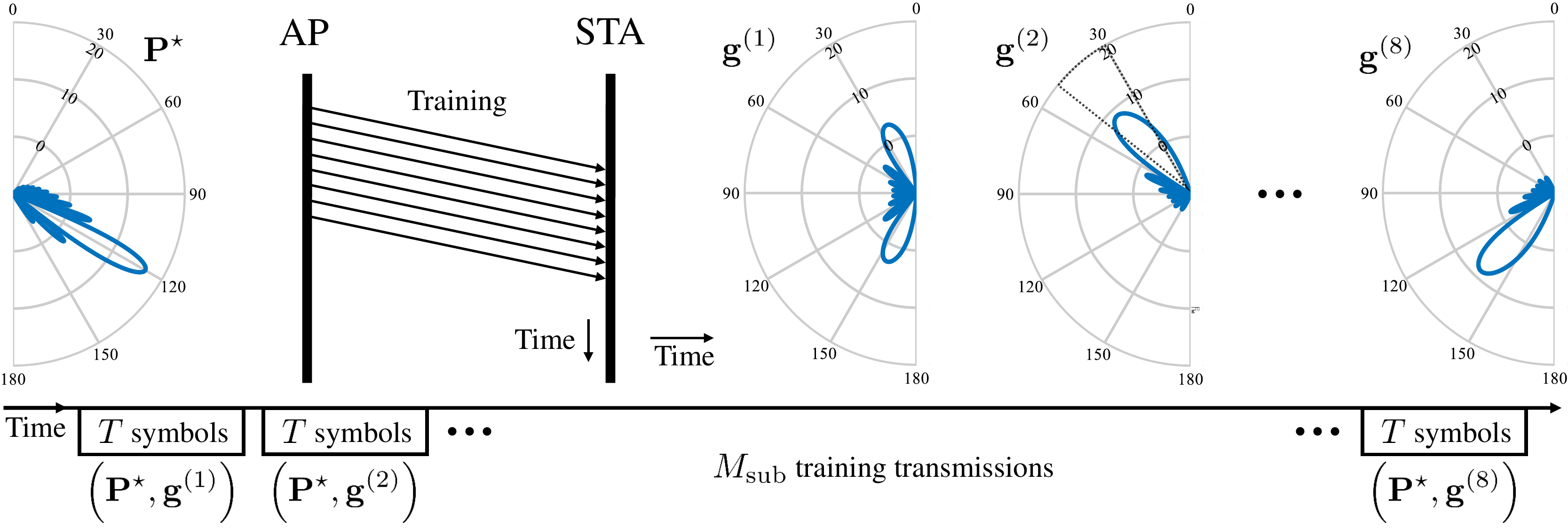}
\caption{Beam selection procedure - Second stage (downlink). The AP uses the optimal beamformer found in stage 1 to transmit training signals. The STA finds its best sector beamformer by sweeping through beamformers in $\Gset_s$. In this example, $M_{\ap} = M_{\ue} = 16$ antennas, $M_{\sub}=8$ antennas, and $N_{\rf}=4$ chains. The sector where $\g^{(2)}$ provides maximum received power (under a single-path model) is highlighted.}
\label{fig_stage2}
\end{figure*}

\item After trying {all possible} AP/user sector beam combinations, the AP obtains the beamformers that {maximize }the received power across subcarriers as
\begin{IEEEeqnarray}{rCl}
\label{eq_optimstage1}
\hspace*{-24pt} \left\{ \PP^{(m_\star)} , \g^{(m'_\star)} , n_\star \right\} = \arg \max_{ \substack{ \PP^{(m)} \in \Pset_s \\ \g^{(m')} \in \Gset_s \\ n \in \{1,\ldots,N_{\rf} \}}} \left\Vert \vbh_{n}^{m,m'} \right\Vert^2_2,
\end{IEEEeqnarray}
which is equivalent to an exhaustive search over $\Bset(M_{\ap}) \times \Gset_s$, given that all of the elements in $\Bset(M_{\ap})$ are used in $\Pset_s$. The optimal \emph{narrow} AP beamformer solving (\ref{eq_optimbeamselection2}) is in column (RF chain) $n_\star$ within $\PP^{(m_\star)}$. Due to the codebook structure in (\ref{eq_orthogonalmatrix}) and (\ref{eq_codebookP1}), this search leads to
\begin{IEEEeqnarray}{rCl}
\label{eq_algorithmsolutionAP}
\p^\star = \bb_{l_{n_\star}(m_\star)}(M_{\ap}).
\end{IEEEeqnarray}
Note that multiple RF chains are used to reduce the number of required training transmissions by testing simultaneously $N_{\rf}$ distinct beamformers using the codebook $\Pset_s$.

\item We use the notation $\PP^\star = \PP^{(m_\star,n_\star)}$, which is the beamforming matrix with $\p^\star$ in all its RF chains.

\item A total of $\tfrac{M_{\ap}}{N_{\rf}} \times M_{\ue}$ training transmissions are required in this stage.

\end{enumerate}

%%%%%%%%%%%%%%%%%%%%%%%%%%%%%%%%%%%%%%%%%%%%%%%%%%%%%%%%%%%%%%%%%%%%%%%%%%%%%%%%%%%%%%%%%%%%%%%%%
%%%%%%%%%%%%%%%%%%%%%%%%%%%%%%%%%%%%%%%%%%%%%%%%%%%%%%%%%%%%%%%%%%%%%%%%%%%%%%%%%%%%%%%%%%%%%%%%%
%%%%%%%%%%%%%%%%%%%%%%%%%%%%%%%%%%%%%%%%%%%%%%%%%%%%%%%%%%%%%%%%%%%%%%%%%%%%%%%%%%%%%%%%%%%%%%%%%

\subsection{Beam Selection - Stage 2 (Downlink)}

\begin{algorithm}[t]
\caption{Analog Beam Selection - Stage 2 (Downlink)}
\begin{algorithmic}[1]
\label{alg_beamselection2}
\renewcommand{\algorithmicrequire}{\textbf{Input:}}
\renewcommand{\algorithmicensure}{\textbf{Output:}}
\REQUIRE $\PP^\star$ known at the AP. $\Gset_s$ known at the STA. Training signal $\x[k]$.
\ENSURE Beamformer $\g^{\left(m'_{\star}\right)}$ known at the STA.
\STATE AP: set $\PP_{\an} = \PP^\star$
\FOR {$m' = 1$ to $\tfrac{M_{\sub}}{N_{\sub}}$}
	\STATE AP: transmit $\x[k]$.
	\STATE STA: set $\g = \g^{(m')}$ and receive the signal.	
	\FOR {every $k \in  \Kset$}
		\STATE STA: estimate $w_{m'}[k]$ using (\ref{eq_stage2channelestimate}).
	\ENDFOR
\ENDFOR
\STATE STA: obtain $\g^{\left(m'_{\star}\right)}$ from (\ref{eq_optimstage2}).
\RETURN $\g^{\left(m'_{\star}\right)}$ known at the STA.
\end{algorithmic} 
\end{algorithm}

In this stage, summarized in Algorithm \ref{alg_beamselection2}, the STA finds the \emph{sector} beam $\g^{m'_\star}$ that maximizes the estimated downlink received sum-power across subcarriers. Fig. \ref{fig_stage2} shows an implementation example. The procedure is the following:

\begin{enumerate}
\item The AP sends training signals using $\PP_{\an} = \PP^\star$. 

\item The STA receives the training signals by sequentially using all the sector beams in $\Gset_s$.

\item Using (\ref{eq_downlinksignal}) with $\g = \g^{\left(m'\right)}$ and $\PP_{\an} = \PP^\star$, the received signal in this downlink is $\y_{\dl}^{\left(m'\right)}[k]$. The corresponding channel coefficient is $w_{m'}[k]$.

\item Following the same procedure as in {Stage 1,} the STA obtains an ML downlink channel coefficient estimate as
\begin{IEEEeqnarray}{rCl}
\label{eq_stage2channelestimate}
\wh_{m'}[k] =\frac{\y_{\dl}^{\left(m'\right)}[k] \x^H[k]}{\Vert \x[k] \Vert_2^2}.
\end{IEEEeqnarray}
We define the vector of downlink channel estimates for STA beamformer $m'$ as $\wbh_{m'} = \left[ \wh_{m'}[1], \ldots , \wh_{m'}[K] \right]^T$.

\item The STA finds its best \emph{sector} beamformer as
\begin{IEEEeqnarray}{rCl}
\label{eq_optimstage2}
\g^{\left(m'_{\star}\right)} = \arg \max_{ \g^{(m')} \in \Gset_s } \left\Vert \wbh_{m'} \right\Vert^2_2,
\end{IEEEeqnarray}
this is, the STA selects the beam that maximizes the total estimated sum-power across subcarriers using exhaustive search within $\Gset_s$ for a fixed AP beamformer.

\item The number of required training sequence transmissions for this stage is $M_{\sub}$.
\end{enumerate}

%%%%%%%%%%%%%%%%%%%%%%%%%%%%%%%%%%%%%%%%%%%%%%%%%%%%%%%%%%%%%%%%%%%%%%%%%%%%%%%%%%%%%%%%%%%%%%%%%
%%%%%%%%%%%%%%%%%%%%%%%%%%%%%%%%%%%%%%%%%%%%%%%%%%%%%%%%%%%%%%%%%%%%%%%%%%%%%%%%%%%%%%%%%%%%%%%%%
%%%%%%%%%%%%%%%%%%%%%%%%%%%%%%%%%%%%%%%%%%%%%%%%%%%%%%%%%%%%%%%%%%%%%%%%%%%%%%%%%%%%%%%%%%%%%%%%%

\subsection{Beam Selection - Stage 3 (Downlink)}

\begin{figure*}[!h]
\centering
\includegraphics[width=1.6\columnwidth]{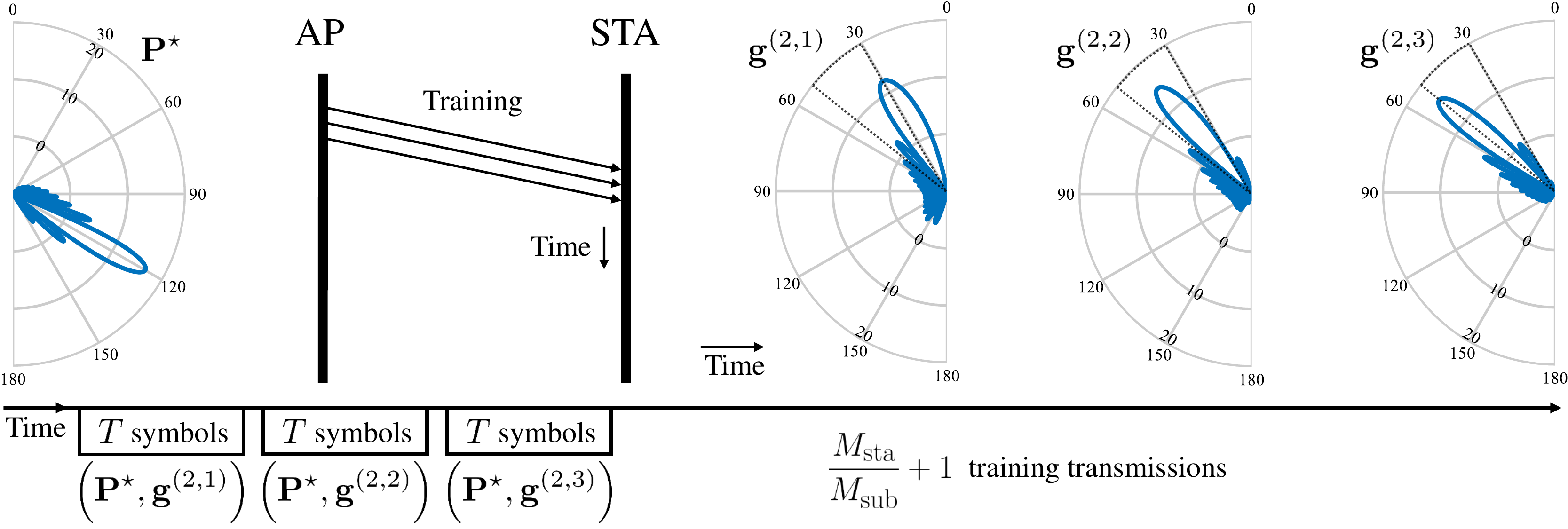}
\caption{Beam selection procedure - Third stage (downlink). The AP uses the optimal beamformer found in stage 1 to transmit training signals. The STA finds its best narrow beamformer by sweeping through beamformers in $\left\{ \g^{(m_\star,n)} \right\}$, i.e. narrow beams that overlap with the optimal sector found in stage 2 (highlighted). In this example, $M_{\ap} = M_{\ue} = 16$ antennas, $M_{\sub}=8$ antennas, and $N_{\rf}=4$ chains.}
\label{fig_stage3}
\end{figure*}

In this stage, summarized in Algorithm \ref{alg_beamselection3}, the STA obtains a \emph{narrow} beam that maximizes the estimated downlink sum-power. Fig. \ref{fig_stage3} shows an implementation example. The procedure is the following.

\begin{enumerate}

\item The AP sends training signals using $\PP_{\an} = \PP^\star$. 

\item The STA receives the training signal by sequentially using all the narrow beams that overlap the sector generated by $\g^{(m'_\star)}$, i.e., the set $\left\{ \g^{(m'_\star , n')} \right\} \subset \Gset$, $n=1,\ldots,\frac{M_{\ue}}{M_{\sub}}+1$.

\item The received signal in this downlink is denoted as $\y_{\dl}^{\left(m'_\star , n' \right)}[k]$, which has the form in (\ref{eq_downlinksignal}) with $\g = \g^{\left(m'_\star , n' \right)}$ and $\PP_{\an} = \PP^\star$. The corresponding channel coefficient is $w_{m'_\star,n'}[k]$.

\item Following the same procedure as in Stages 1 and 2, the STA obtains an ML downlink channel coefficient estimate denoted $\wh_{m'_\star , n' }[k]$. The vector of downlink channel estimates for all subcarriers is $\wbh_{m'_\star,n'} = \left[ \wh_{m'_\star,n'}[1], \ldots , \wh_{m'_\star,n'}[K] \right]^T$.

\item The terminal finds its best narrow beamformer as
\begin{IEEEeqnarray}{rCl}
\label{eq_optimstage3}
\g^\star = \arg \max_{ \g^{ \left( m'_\star , n' \right) } \in \Gset } \left\Vert \wbh_{m'_\star , n'} \right\Vert^2_2.
\end{IEEEeqnarray}
Note that this is an heuristic algorithm since it assumes that a solution for $\g^\star$ found using (\ref{eq_optimstage2}) and (\ref{eq_optimstage3}) is equivalent to an exhaustive search in (\ref{eq_optimbeamselection2}). {This assumption} is only true when the channel has one path. When multiple paths exist, our algorithm might converge to a suboptimal solution even at high SNR. However, the performance loss due to this misalignment is negligible in the scenarios analyzed in Section \ref{sec_results}.

\item The number of required training sequence transmissions for this stage is $\frac{M_{\ap}}{M_{\sub}}+1$.

\end{enumerate}

At this point in the algorithm, both the AP and the STA know the beamformers that maximize the received power. %However, estimation on Stage 1 was made using sector UE beamformers with lower gain than the one provided by $\g^\star$. Thus, a second uplink stage using $\g=\g^\star$ at the UE can improve the estimation and reduce beam selection errors. We found that an additional uplink stage provides negligible performance gains in the scenarios presented in Section \ref{sec_results}.

\begin{algorithm}[!t]
\caption{Analog Beam Selection - Stage 3 (Downlink)}
\begin{algorithmic}[1]
\label{alg_beamselection3}
\renewcommand{\algorithmicrequire}{\textbf{Input:}}
\renewcommand{\algorithmicensure}{\textbf{Output:}}
\REQUIRE $\PP^\star$ known at the AP. $\g^{\left(m'_{\star}\right)}$ and $\Gset$ known at the STA. Training signal $\x[k]$.
\ENSURE Beamformer $\g^\star$ known at the terminal.
\STATE AP: set $\PP_{\an} = \PP^\star$.
\FOR {$n'=1$ to $\tfrac{M_{\ap}}{M_{\sub}}+1$}
	\STATE AP: transmit $\x[k]$.
	\STATE STA: set $\g= \g^{(m'_{\star},n')}$ and receive the signal.
	\FOR {every $k \in \Kset$}
		\STATE STA: estimate $w_{m'_\star , n'}[k]$.
	\ENDFOR
\ENDFOR
\STATE STA: find $\g^\star$ using (\ref{eq_optimstage3}).
\RETURN $\g^\star$ known at the STA.
\end{algorithmic} 
\end{algorithm}

\subsection{Beam Selection - Stage 4 (Analog Beamforming Matrix Construction)}
Let $\p_u^\star$ and $\g_u^\star$ denote the optimum AP and STA beamformers for user $u$, respectively, obtained by applying the previous stages to all the STAs. We construct the analog beamforming matrix for the multiuser downlink in (\ref{eq_downlinksignal}) depending on the number of RF chains and the number of users:
\begin{itemize}
\item If $U=N_{\rf}${, }the analog beamforming matrix is constructed as $\PP_{\an} = \left[ \p_1^\star \ \p_2^\star \cdots \p_U^\star \right]$. Hence, every RF chain points to a different STA.
\item If $U < N_{\rf}$, the beamforming vector for any given user should appear at least once as a column in the analog beamforming matrix. For example, if $N_{\rf}=4$ and $U=2$, $\PP_{\an} = \left[ \p_1^\star \  \p_1^\star \  \p_2^\star \  \p_2^\star \right]$ is an acceptable matrix. The number of times that a beamforming vector appears in the matrix has no influence in our algorithm\footnote{Further algorithm extensions could consider power constraints per RF chain. In that case, the multiplicity of a vector in the analog beamforming matrix might be used as a design parameter.}.
\item If $U > N_{\rf}$, simultaneous communications with all the STAs are not feasible using linear precoding only since $\rank \left( \PP_{\an} \PP_{\di}[k] \right) < U$. This is also the case when two or more STAs share the same AP beamformer vector. In such cases, other multiple access techniques should be combined with hybrid precoding to communicate with all the users. The algorithm could also be repeated for some STAs to find alternative beams to form feasible analog beamforming matrices.
\end{itemize}

\begin{table*}[!t]
\caption{Training Overhead and Computational Cost}%
\label{tab_complexity}
\centering
\begin{tabular}{cccc}
\hline \hline
\textbf{Algorithm}		&	Training Overhead	&	Single-User Complexity	&	Multi-User Complexity	\\
\hline
This work				&	$\frac{M_{\ap}}{N_{\rf}}M_{\sub}+M_{\sub}+\frac{M_{\ue}}{M_{\sub}}+1$	&	$\mathcal{O}\left(K_{\tx}^2\frac{M_{\ap}}{N_{\rf}}M_{\sub}\right)$*	&	$\mathcal{O}\left(K_{\tx}^2\frac{M_{\ap}}{N_{\rf}} U\right)$	\\
\cite{venugopal2017}	&	40 - 100**	&	-	&	-	\\
\cite{alkhateeb2016}	&	-			&	$\mathcal{O}\left(K M_{\ap}^2 M_{\ue}+KN_{\rf}^3\right)$***	&	-	\\
\cite{sohrabi2017}		&	-			&	-	&	$\mathcal{O}\left(K M_{\ap}^2 U\right)$	\\
\hline \hline
\multicolumn{4}{l}{* $K_{\tx}^2$ is in the order of $K$, for example, $K_{\tx}=16$ and $K=512$.} \\
\multicolumn{4}{l}{** Values reported in \cite{venugopal2017} for 16 to 32 antennas at the terminals (similar to this work).} \\
\multicolumn{4}{l}{*** Obtained from \cite[Algorithm 2]{alkhateeb2016}.} \\
\hline \hline
\end{tabular}
\end{table*}

\subsection{Digital Beamformer Design}
After obtaining a suitable $\PP_{\an}$, the AP calculates a digital beamformer $\PP_{\di}[k]$. We assume that the analog precoding matrix is able to resolve signals to different users such that $\rank \left( \PP_{\an}[k] \right) = U$, and we use block-diagonalization (BD) to eliminate residual inter-user interference after analog beamforming \cite{spencer2004}. The first step is to obtain equivalent MISO channel estimates (as given by (\ref{eq_equivalentchannel})) using the following procedure, which is executed for every STA:
\begin{itemize}
\item User $u$ sends the training signal in the uplink using $\g_u^\star$.
\item The AP uses $\PP_{\an}$ as constructed from the beam selection procedure.
\item The AP obtains a maximum likelihood channel coefficient estimation at each RF chain. Then the equivalent MISO channel estimate for user $u$ is $\hbh_{\eq,u}[k] = \left[ \vh_{u,1}[k], \ldots, \vh_{u,N_{\rf}}[k] \right]$, where $\vh_{u,n}[k]$ denotes the estimate for RF chain $n$.
\item $U$ total training transmissions are required in this stage.
\end{itemize}
BD enforces the zero-interference constraints, i.e.
\begin{IEEEeqnarray}{rCl}
\hbh_{\eq,u}[k] \left[ \PP_{\di}[k] \right]_{:,u'} = 0, \ \forall u \neq u', \ \forall k,
\end{IEEEeqnarray}
and then maximizes the received signal power at each user, as shown in \cite{spencer2004}.

%% file: Results.tex
\section{Numerical Results and Discussion}
\label{sec_results}

In this section, we describe numerical experiments that validate the proposed algorithm under standard operating conditions in indoor environments. Our algorithm is designed for systems having the 6 features described in Section \ref{sec_intro}. {To the best of our knowledge, there are no other algorithms that incorporate all of these features and simpler systems found in literature can be regarded as special cases of the general system considered here. Because of this, we analyze three specific wideband mmWave scenarios for which there are other algorithms available and comparison is possible:}
\begin{enumerate}[label=S\arabic*.]
\item A single-user system with hybrid beamforming at the AP and analog beamforming (multiple antennas connected to one RF chain) at the STA. In this case, we compare our algorithm with the approaches in \cite{alkhateeb2016} and \cite{sohrabi2017}.

\item A multiuser system with hybrid beamforming at the AP and single-antenna STAs. We compare our algorithm with the work in \cite{sohrabi2017}.

\item A system with the 6 features described in Section \ref{sec_intro}. We compare our algorithm with ideal fully-digital BD precoding \cite{spencer2004}, {which would be costly in terms of hardware and would require one RF chain per antenna.} %As mentioned before, there are no other algorithms for this general scenario.
\end{enumerate}
{We start by analyzing the training overhead and computational complexity of the algorithms}, and then analyze their sum-rate (spectral efficiency) performance.

\subsection{Training Overhead}
Since the approaches in \cite{alkhateeb2016} and \cite{sohrabi2017} require channel matrix knowledge, they need an additional wideband channel estimation algorithm. We use the frequency domain method proposed in \cite{venugopal2017} to obtain the channel matrix estimate. Our algorithm requires $\frac{M_{\ap}}{N_{\rf}}M_{\sub}+M_{\sub}+\frac{M_{\ue}}{M_{\sub}}+1$ training transmissions to converge to a beamforming solution. In the practical scenarios described later in this section (terminals with 16 or 32 antennas), this value is approximately 40 - 80 training transmissions compared with the 40 - 100 transmissions to estimate the channel matrix using \cite{venugopal2017}. Thus, our algorithm is similar to other state-of-the-art methods with respect to training overhead.

%In the first case, we compare the sum-rate performance of our algorithm versus the approaches in \cite{alkhateeb2016} and \cite{sohrabi2017} and, in the second case, we use \cite{sohrabi2017} for comparison. Furthermore, we use ideal fully-digital beamforming as benchmark (using the channel singular vectors as beamformers or the BD precoder in the multiuser case), which would require one RF chain per antenna at every terminal.

\subsection{Computational Complexity}
\label{sec_complexity}
Regarding computational complexity, our algorithm calculates $K_{\tx}$ inner products for each training transmission, stores them on a vector, and then compares its magnitude to that of vectors obtained with other beamformers in each stage. The process is repeated for every user. Since the number of training transmissions in the first stage is much greater than the number required in the remaining stages, the dominant term in the computational cost of our algorithm is $\mathcal{O}\left(K_{\tx}^2\frac{M_{\ap}}{N_{\rf}} M_{\sub} U\right)$. Note that this cost depends on the squared number of pilot subcarriers $K_{\tx}^2$, which is typically in the order of the total number of subcarriers $K$, and is linear with respect to the number of antennas at the AP (e.g., $K_{\tx}=16$ and $K=512$). However, the complexity is linear with respect to the number of antennas in the STA subarray (not the full array) and it depends on $N_{\rf}^{-1}$. Thus, our algorithm effectively harnesses multiple RF chains at the AP and subarrays at the STAs to reduce complexity. Table \ref{tab_complexity} shows the training overhead and complexity comparison of our algorithm with \cite{venugopal2017}, \cite{alkhateeb2016} and \cite{sohrabi2017} in the special cases mentioned above. We obtained the computational cost from the number of operations required to execute each algorithm (e.g. inversion of an $n \times n$ matrix has a cost of $\mathcal{O}(n^3)$). Note that current algorithms have complexities depending on $M_{\ap}^2$ and $N_{\rf}^3$. Hence, our algorithm has a lower computational burden with respect to current methods, taking advantage of hardware architectures to reduce complexity.

%As an example, for a single-user wideband system, \cite[Alg. 2]{alkhateeb2016} has a computational cost of $\mathcal{O}\left(K M_{\ap}^2 M_{\ue}+KN_{\rf}^3\right)$ (the factor $KN_{\rf}^3$ comes from matrix inversions) compared to our algorithm's $\mathcal{O}\left(K_{\tx}^2\frac{M_{\ap}}{N_{\rf}}M_{\ue}\right)$. Similarly, in a multiuser system with single-antenna STAs, the algorithm in \cite{sohrabi2017} has a computational cost of $\mathcal{O}\left(K M_{\ap}^2 U\right)$ whereas our algorithm has a cost of $\mathcal{O}\left(K_{\tx}^2\frac{M_{\ap}}{N_{\rf}} U\right)$.

\subsection{Statistical Algorithm Characterization}
We provide an algorithm performance characterization under a statistical single-path channel model with angles of departure and arrival uniformly distributed over the interval $[0,\pi]$. This is represented as $L=1$ and $\alpha_1 \sim \cgaussian(0,1)$ in (\ref{eq_channelmodel}), for which we also used a $-20$ dB coupling between adjacent antenna elements as described in Section \ref{sec_sytemmodel}. Under these conditions, the channel power constraint is $\E \left\{ \Vert \HH[k] \Vert_2^2 \right\} = M_{\ap} M_{\ue}$, $\forall k$. ULAs with 16 and 32 antennas were analyzed since they fit typical form factors for indoor terminals. Other parameters are summarized in Table \ref{tab_parameters} and were taken from \cite[p. 446]{iso2014}.
\begin{table}[!t]
\caption{Simulation Parameters}%
\label{tab_parameters}
\centering
\begin{tabular}{lr}
\hline \hline
\textbf{Parameter}										&	\textbf{Value}	\\
\hline
Reference frequency, $f_0$								&	60 GHz			\\
Channel central frequency								&	58.32 GHz		\\
OFDM sampling rate										&	2640 MHz		\\
OFDM sampling time										&	0.38 ns			\\
Subcarrier spacing										&	5156.25 KHz		\\
Normalized element spacing, $d$							&	0.5				\\
Number of subcarriers, $K$								&	512				\\
Number of pilot subcarriers, $K_{\tx}$					&	4, 16 and 64	\\
Number of antennas at the AP, $M_{\ap}$					&	16 and 32		\\
Number of antennas at the STAs, $M_{\ue}$				&	16 and 32 		\\
Number of subarray antennas at the STAs, $M_{\sub}$		&	8				\\
Number of RF chains at the AP, $N_{\rf}$				&	4				\\
Number of training symbols, $T$							&	64				\\
\hline \hline
\end{tabular}
\end{table}

\begin{figure}
\subfloat[]{\includegraphics[width=\columnwidth]{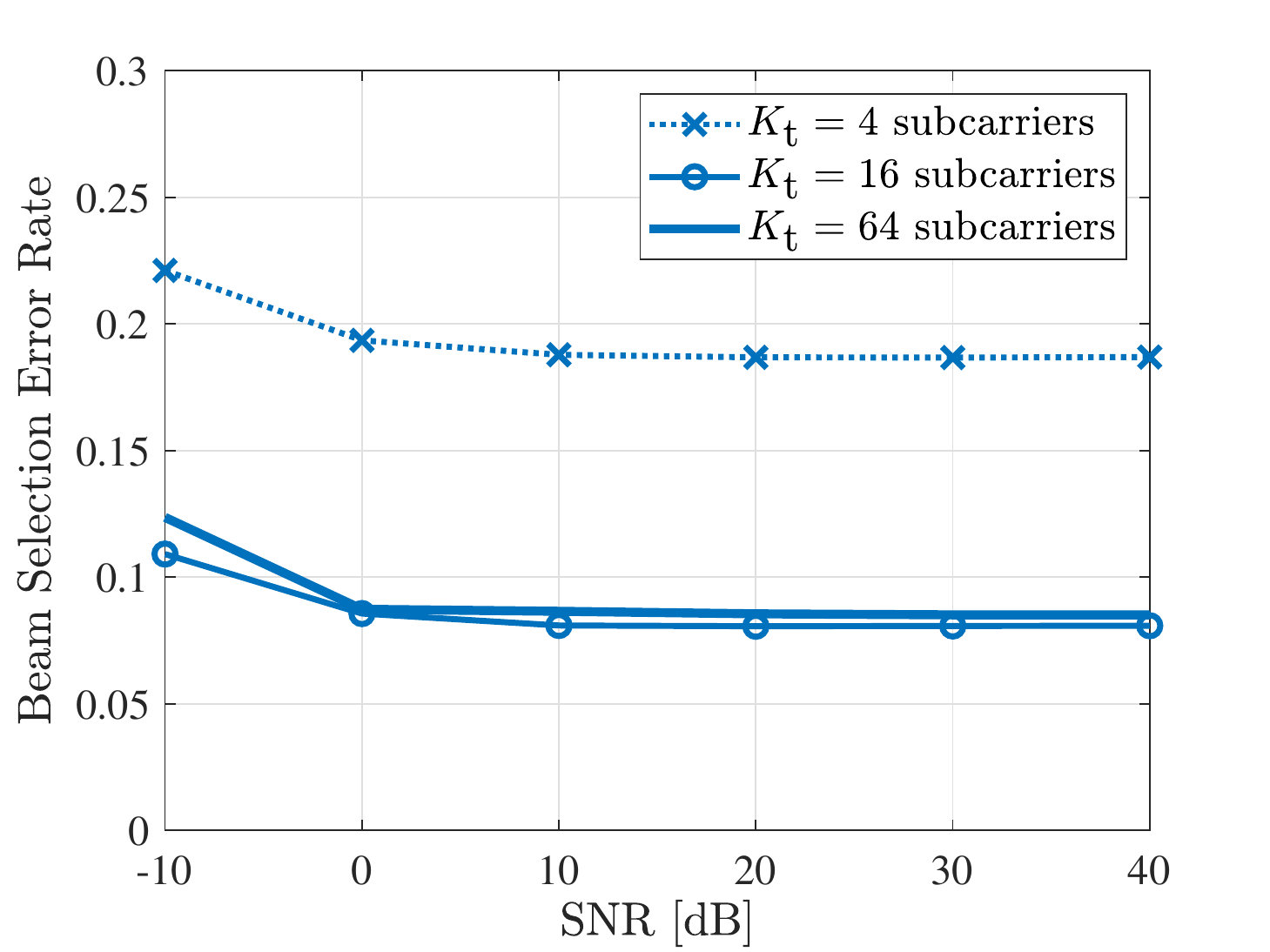}\label{fig_bser}}\\
\subfloat[]{\includegraphics[width=\columnwidth]{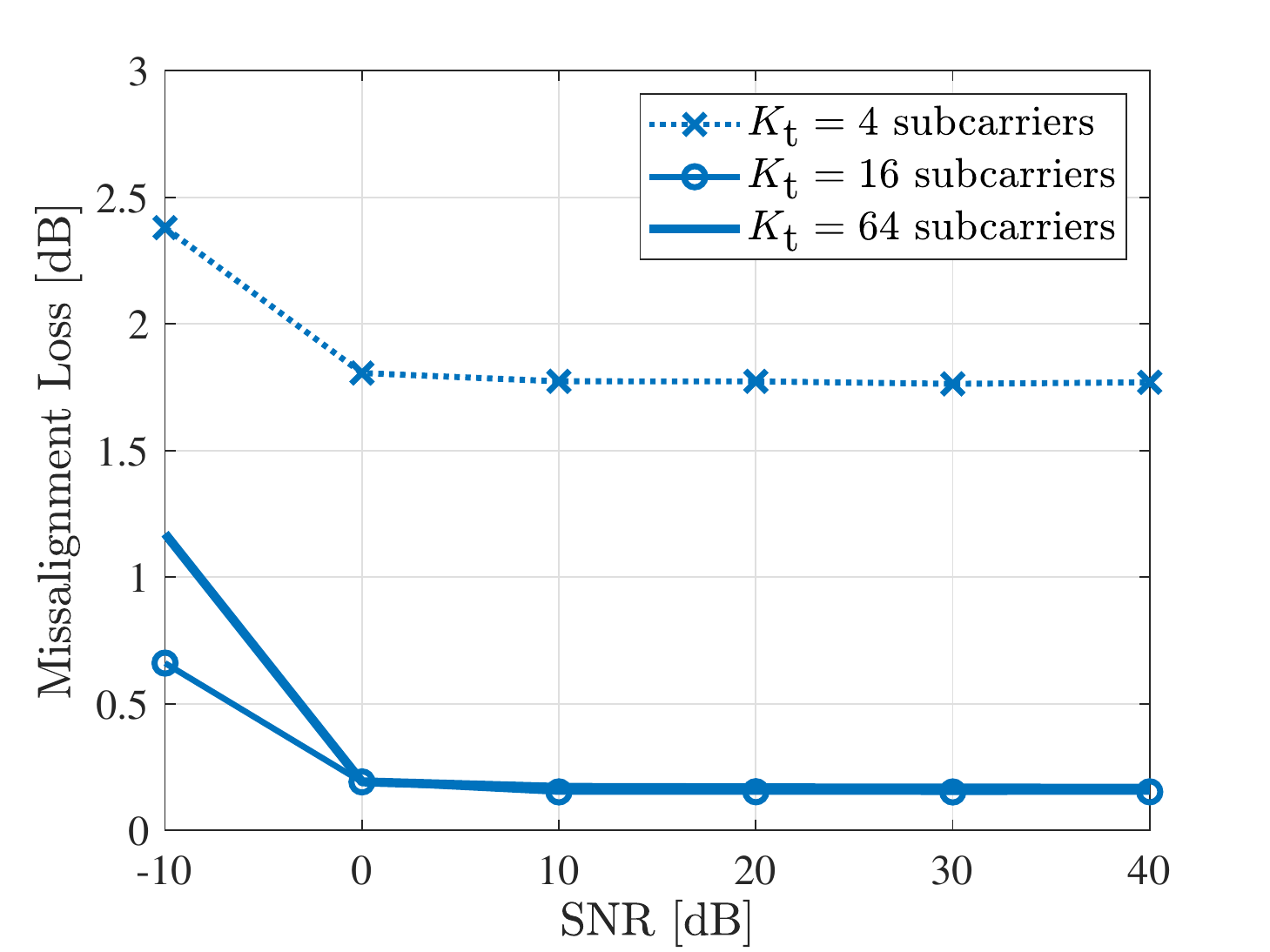}\label{fig_losses}}
\caption{{(a) Beam selection error rate. (b) Missalignment loss. Plots for different number of pilot subcarriers for $M_{\ap} = M_{\ue} = 16$.}}
\end{figure}

{The algorithm's performance is analyzed in terms of three metrics. The first metric is the error rate of finding the global optimum of} (\ref{eq_optimbeamselection1}) {given a large number of channel realizations.} We call this parameter the beam selection error rate (BSER), and we obtained it for $10^5$ channel realizations as shown in Fig. \ref{fig_bser} for different numbers of pilot subcarriers. BSER decays when increasing SNR until approximately 10 dB. After this value, BSER is constant regardless of the number of subcarriers. This BSER bound (approximately 0.08 for $M_{\ap}=M_{\ue}=16$ and 0.19 for $M_{\ap}=M_{\ue}=32$) is due to the element radiation pattern selected for the simulation (see (\ref{eq_radiationpattern})), which has nulls in the array axis directions. Those nulls affect detection given that all beamformers in the codebooks have approximately the same gain for AoD/AoA close to $0$ and $\pi$ radians. We also observe a lower BSER for $K_{\tx}=16$ pilot subcarriers. Using more pilot subcarriers spreads the power across the bandwidth and thus diminishes the channel estimates precision, while less subcarriers fail to capture beam squint. For comparison, we also used a channel model with 3 paths (uniformly distributed AoD and AoA with relative powers 0, $-10$, and $-10$ dB, preserving the channel power constraint). The algorithm has a BSER lower bound at $\text{BSER} \approx 0.07$ for the 3-path channel with $M_{\ap}=M_{\ue}=16$. {The errors occur} mostly when there are 2 dominant paths with similar total sum-power. {BSER gets larger when increasing the number of antennas due to a larger codebook.}

{The second performance metric is the average loss due to missalignment, which is defined as the average power difference (across subcarriers) between the algorithm's beamforming solution and the optimal solution in the codebook. Losses have the same behavior as the BSER, decreasing until the SNR reaches approximately 10 dB as shown in Fig.} \ref{fig_losses}. {Losses are approximately constant for larger SNRs (0.2 dB for $M_{\ap}=M_{\ue}=16$ and 2.4 dB for $M_{\ap}=M_{\ue}=32$ with 16 pilot subcarriers). When the number of antennas increases, losses are larger due to higher BSER.}

%\Figure[t!](topskip=0pt, botskip=0pt, midskip=0pt)[width=0.97\columnwidth]{Figures/BSER.eps}
%{Beam selection error rate for single-path and 3-path channel models with $M_{\ap} = M_{\ue} = 16$. Plots for different number of pilot subcarriers.\label{fig_bser}}
%\begin{figure}
%\centering
%\includegraphics[width=\columnwidth]{Figures/BSER.eps}
%\caption{Beam selection error rate for single-path and 3-path channel models with $M_{\ap} = M_{\ue} = 16$. Plots for different number of pilot subcarriers.}
%\label{fig_bser}
%\end{figure}

The third performance metric is the achievable sum-rate $R$ (spectral efficiency) defined as
\begin{IEEEeqnarray}{cCl}
\label{eq_rate}
R = \frac{1}{K} \sum_{k=1}^K \sum_{u=1}^U \E \left[ \log_2 \left( 1 + \frac{ \eta_u }{ \eta_u^{\inter} + \sigma_z^2 } \right) \right],
\end{IEEEeqnarray}
where $\eta_u$ and $\eta_u^{\inter}$ are the desired signal and the interference powers given by
\begin{IEEEeqnarray}{cCl}
\eta_u = \rho_u \left| \g_u^H \HH_u[k] \PP_{\an} \left[ \PP_{\di}[k] \right]_{:,u} \right|^2,\\
\eta_u^{\inter} = \sum\limits_{u' \neq u} \rho_{u'} \left| \g_u^H \HH_u[k] \PP_{\an} \left[ \PP_{\di}[k] \right]_{:,u'} \right|^2,
\end{IEEEeqnarray}
respectively. BD guarantees that $\eta_u^{\inter}$ is set to zero under perfect CSI. {We assume that OFDM eliminates inter-symbol interference from the system, so that its power is not considered in} (\ref{eq_rate}).  If $\rank \left( \PP_{\an} \right) < U$ for a particular channel realization, the result is not considered for the calculation. We evaluated the achievable sum-rate of our algorithm in the three scenarios above for $10^3$ channel realizations.

\subsubsection{Single-User System}
\begin{figure}
\subfloat[]{\includegraphics[width=\columnwidth]{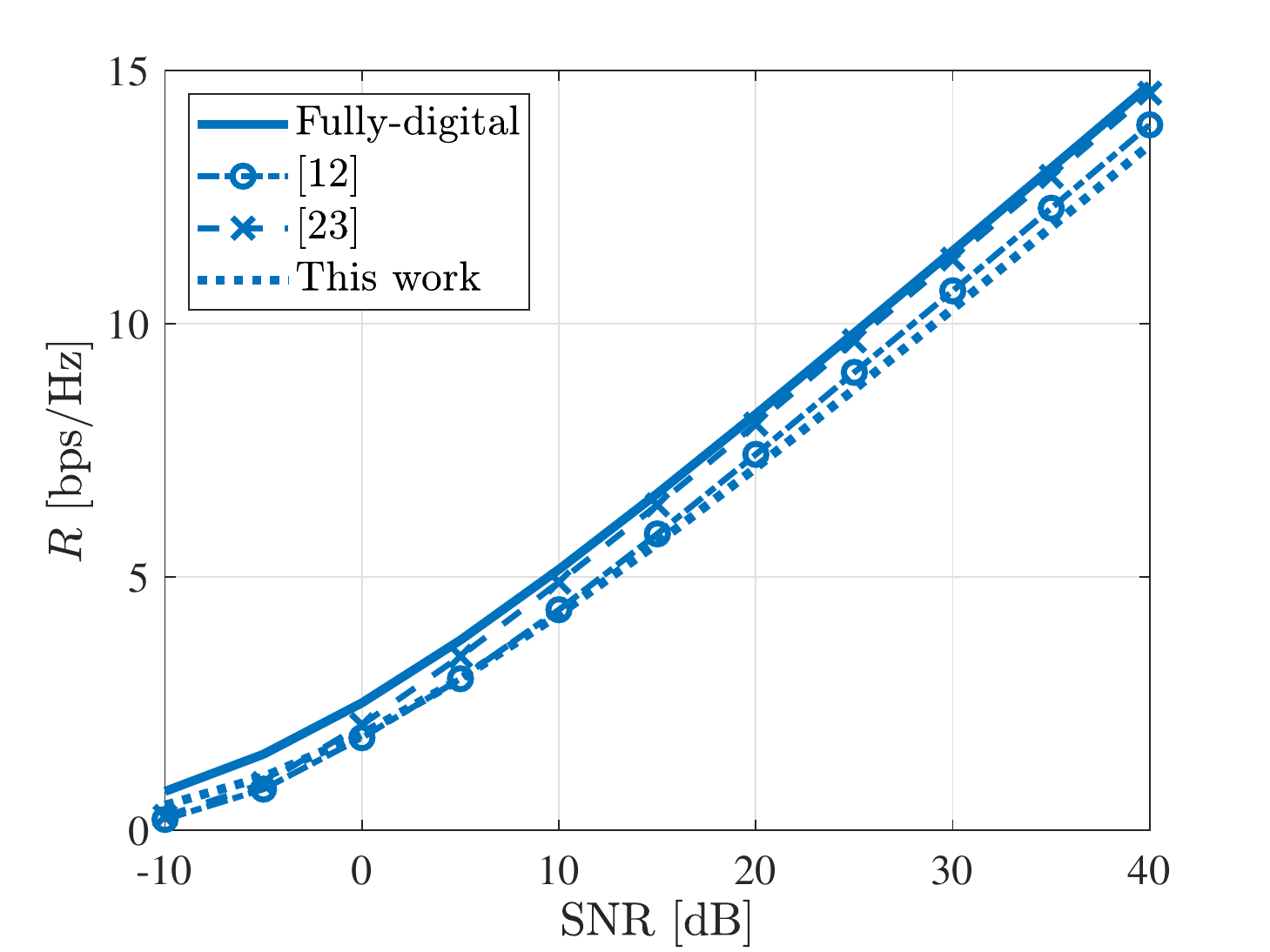}\label{fig_single1path}}\\
\subfloat[]{\includegraphics[width=\columnwidth]{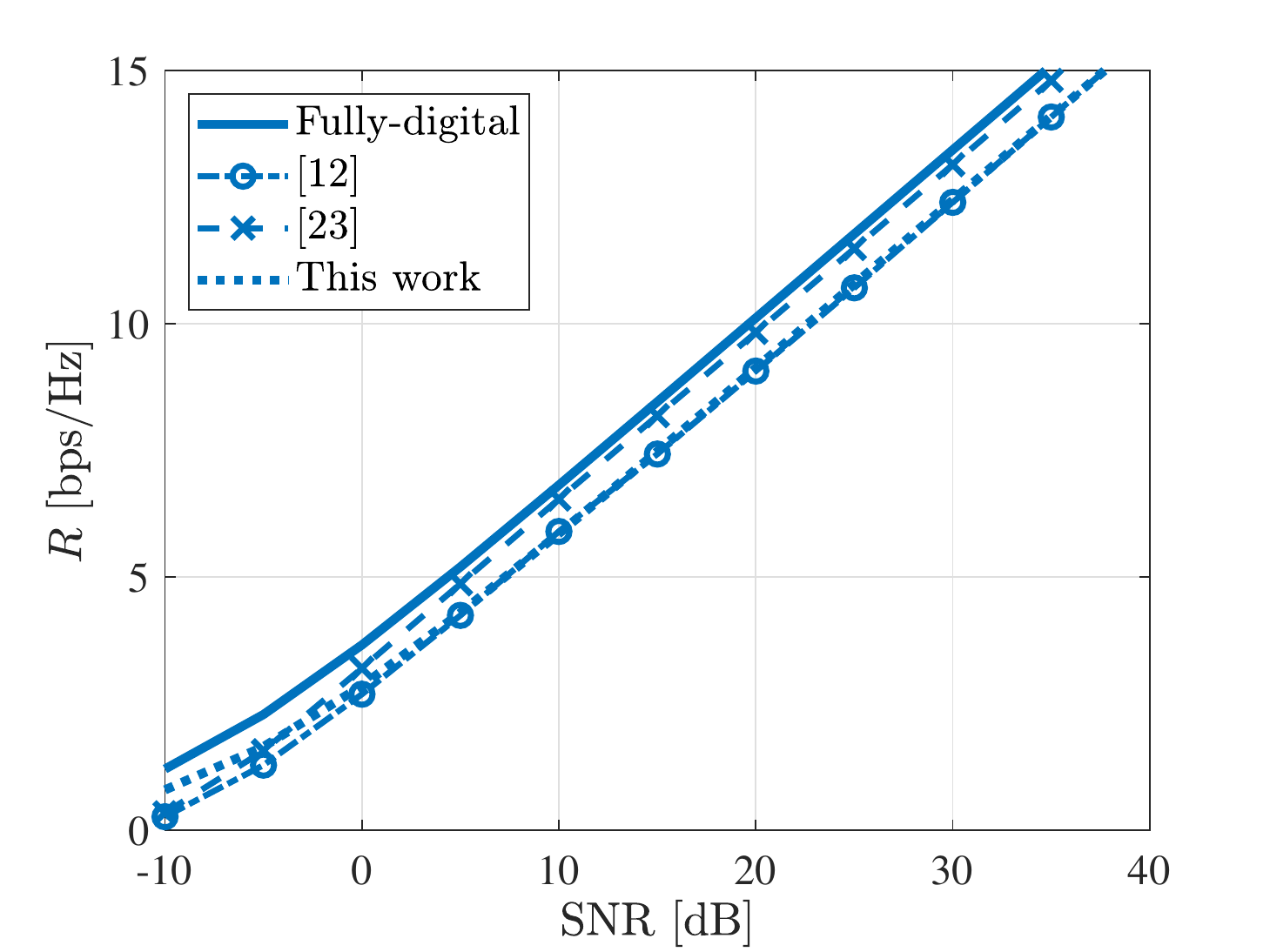}\label{fig_single3path}}
\caption{{Achievable rate in the single-user system for a channel with (a) 1 path, and (b) 3 paths. Plots for different algorithms with $M_{\ap} = M_{\ue} = 32$.}}
\label{fig_singleuserrate}
\end{figure}
In this scenario there is one hybrid-beamforming-capable AP serving one STA that performs only analog beamforming (multiple antennas connected to one RF chain). Since there are no subarrays in this scenario, we replaced stages 2 and 3 of our algorithm with an exhaustive search over $\Bset(M_{\ue})$ (narrow beamformers at the STA). We used the parameters in Table \ref{tab_parameters} and calculated the achievable rate of our algorithm, \cite{alkhateeb2016}, \cite{sohrabi2017}, and ideal fully-digital beamforming (using the singular values of the channel matrix). Furthermore, we used the channel estimation method in \cite{venugopal2017} with the same number of training frames as our algorithm. This estimate was set as an input for the other methods. All of the algorithms used the same power constraints for a fair comparison. Fig. \ref{fig_singleuserrate} shows the achievable rate results in this scenario. Our algorithm performs better than \cite{alkhateeb2016} and \cite{sohrabi2017} below 0 dB. At high SNR and the 3-path channel model, our algorithm performs below \cite{alkhateeb2016} and achieves approximately the same rate as \cite{sohrabi2017} ($\sim3$ dB below the ideal fully-digital beamforming performance). For the 1-path channel model, \cite{sohrabi2017} performs marginally better than our algorithm but the $\sim3$ dB difference with respect to BD is maintained. This difference is due to the codebook design with a fixed number of available beams that does not allow to steer the beam accurately in certain directions. {This (moderate) rate loss in the proposed algorithm is the price to pay for a much lower complexity, as discussed in Section} \ref{sec_complexity}.
\begin{figure}
\subfloat[]{\includegraphics[width=\columnwidth]{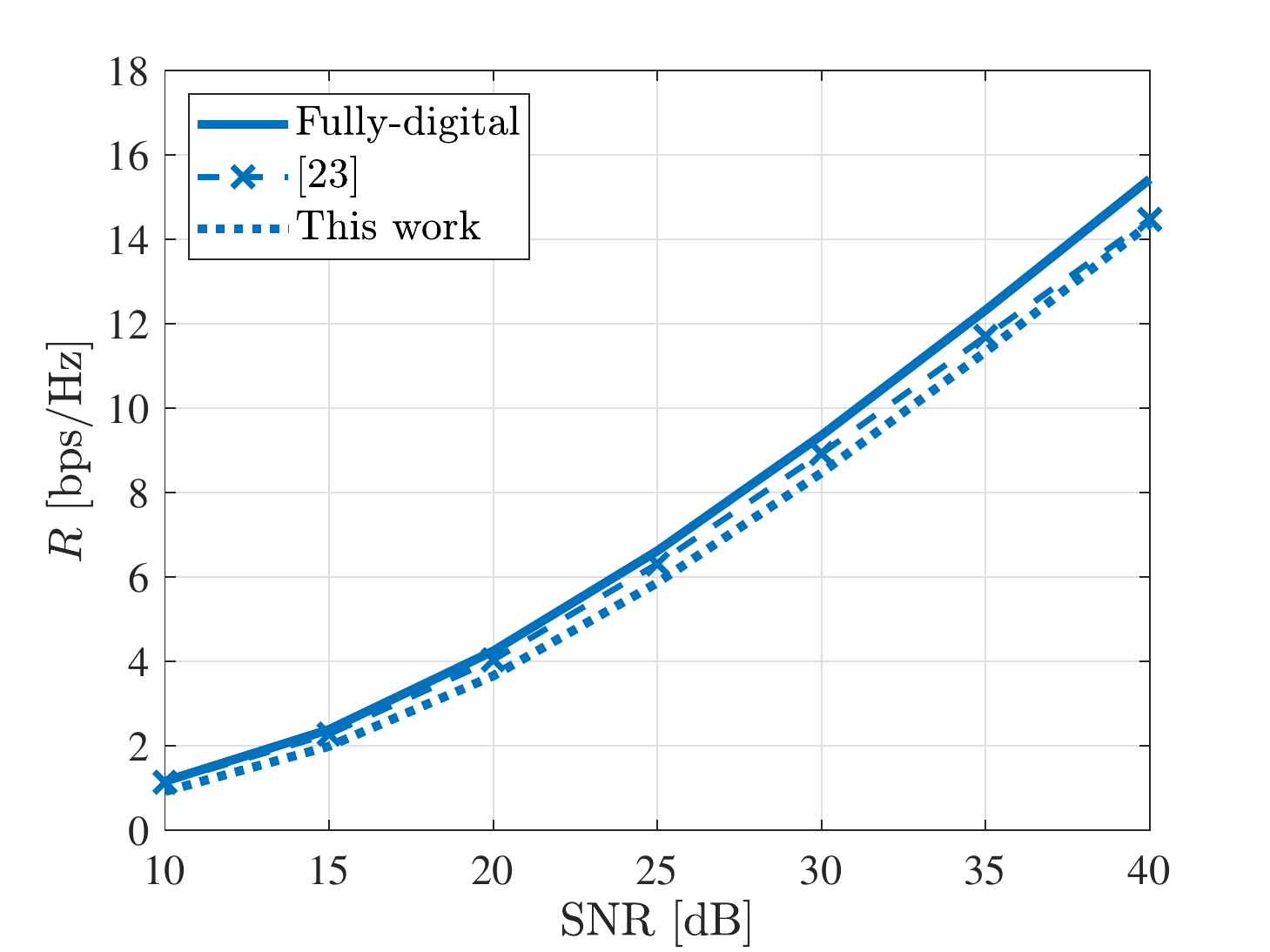}\label{fig_multiuser2u}}\\
\subfloat[]{\includegraphics[width=\columnwidth]{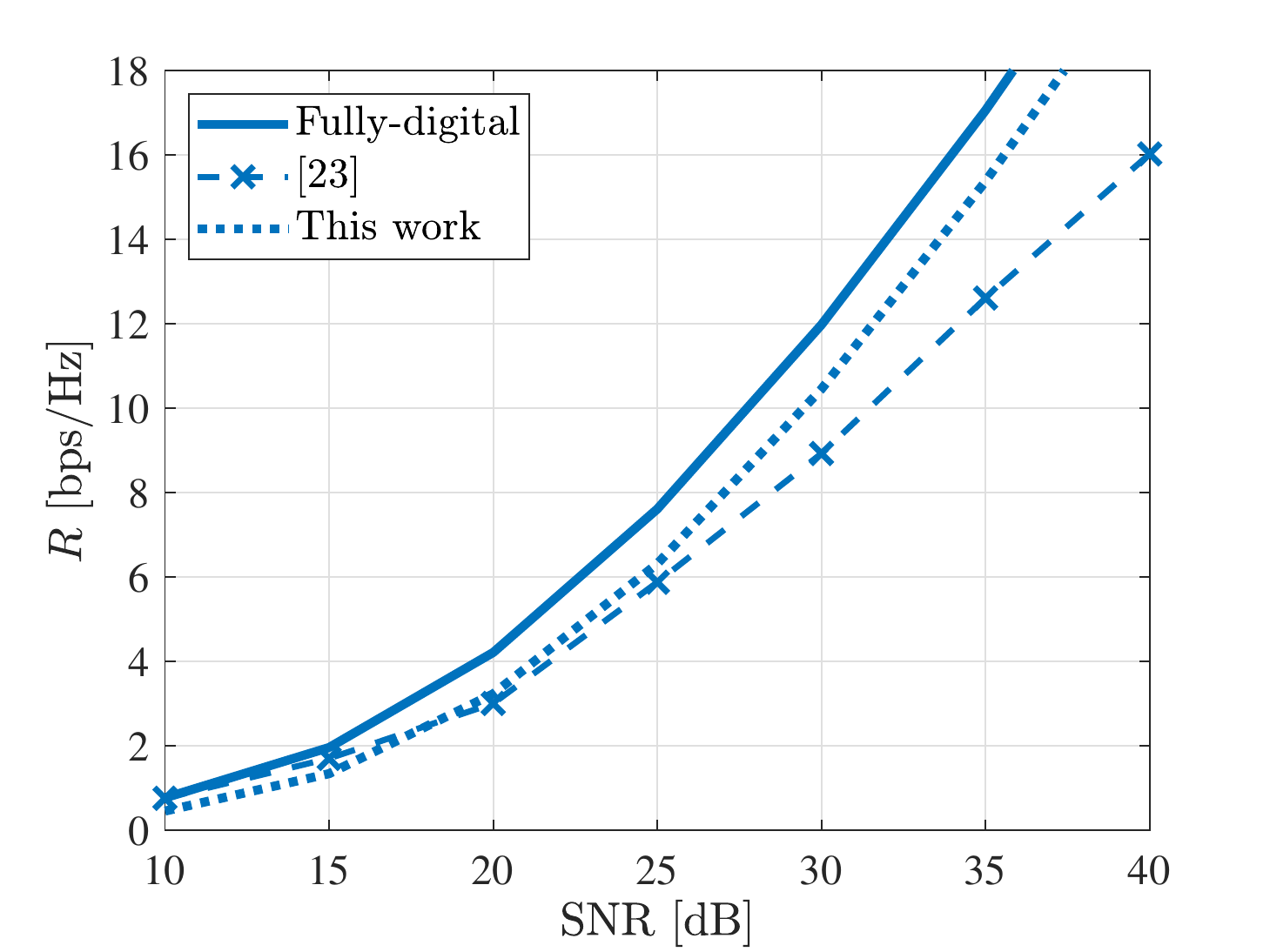}\label{fig_multiuser4u}}
\caption{Achievable sum rate in the multiuser single-antenna STAs system with 32 antennas and 4 RF chains at the AP. (a) 2 users, and (b) 4 users. Our approach requires 8 training transmissions, while \cite{sohrabi2017} requires 512 frames using the channel estimation in \cite{venugopal2017}.}
\label{fig_multiuserrate}
\end{figure}

\subsubsection{Multiuser System with Single-Antenna STAs}
In this scenario, stages 2 and 3 of our algorithm are omitted since there are no arrays at the STAs. Only stage 1 is executed to find the best beamformers at the AP for every user, which requires only $\frac{M_{\ap}}{N_{\rf}}$ training transmissions. We compared our algorithm achievable rate to that of \cite{sohrabi2017}. {However, we found that }\cite{sohrabi2017} {is highly sensitive to channel estimation errors and that }\cite{venugopal2017} {requires more training transmissions than our algorithm to perform well in this scenario. Hence, we used $\frac{M_{\ap}}{N_{\rf}}$ training frames for our algorithm and $16M_{\ap}$ training frames for the channel estimation with} \cite{venugopal2017}. Results are shown in Fig. \ref{fig_multiuserrate} for $M_{\ap}=32$ antennas. We found that the multiuser algorithm in \cite{sohrabi2017} is unfeasible at SNRs below $10$ dB and the results are omitted in those conditions. Our algorithm's performance is $\sim1.5$ dB below fully-digital BD in this scenario. For $U=2$ users, the performance of \cite{sohrabi2017} is marginally better than our algorithm's but the difference decreases until they are the same at 40 dB. For $U=4$ users, our algorithm outperforms \cite{sohrabi2017} for $\snr \geq 20$ dB and it also reduces both training overhead and computational complexity in this scenario. {We note that our approach is not sensitive to channel estimation errors,} which negatively impacts the performance of  \cite{sohrabi2017} at high SNR.

\subsubsection{Multiuser System with Hybrid AP and Subarrays at the STAs}
\begin{figure}
\centering
\subfloat[]{\includegraphics[width=\columnwidth]{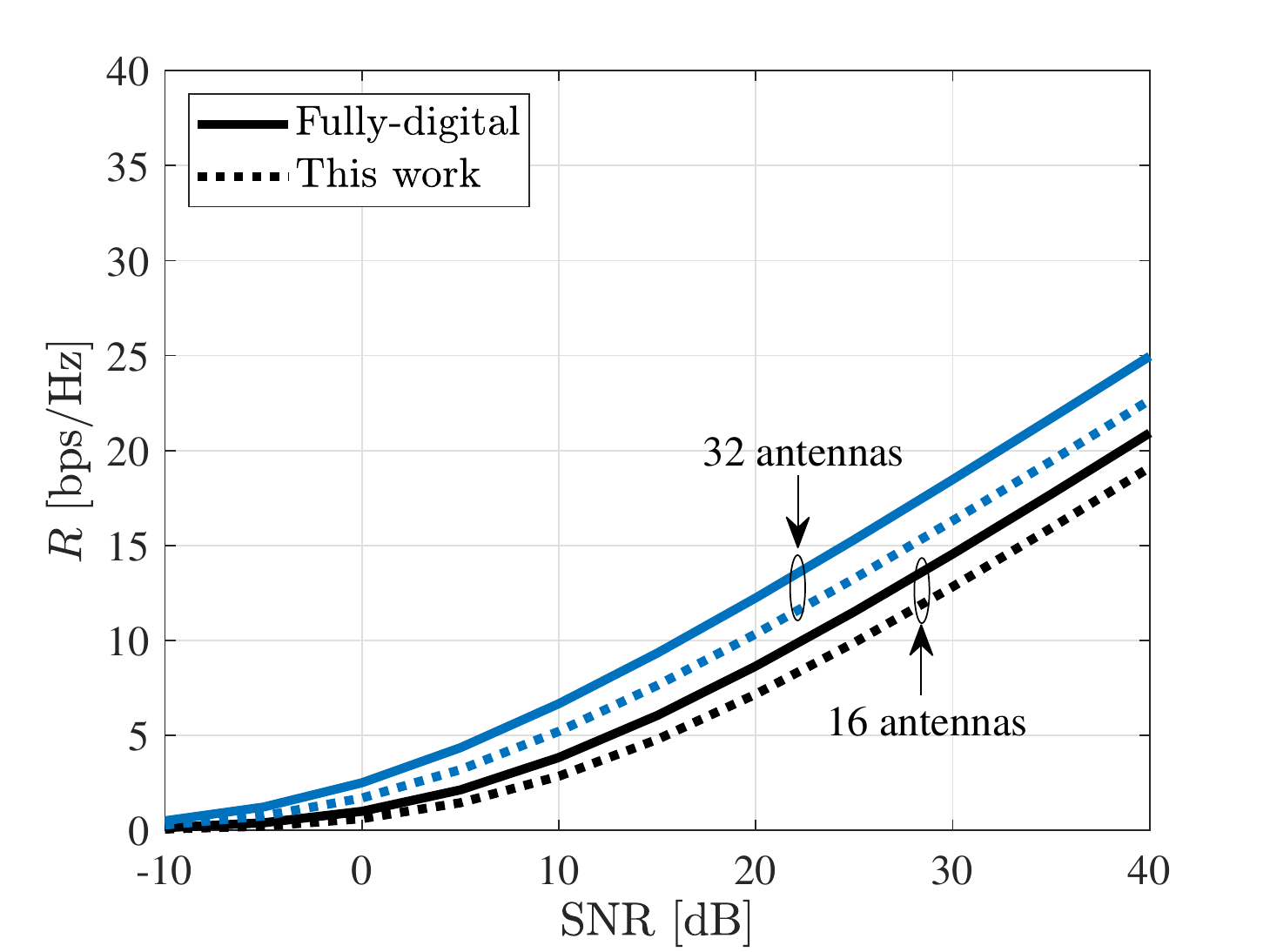}\label{fig_sumrate2users}}\\
\subfloat[]{\includegraphics[width=\columnwidth]{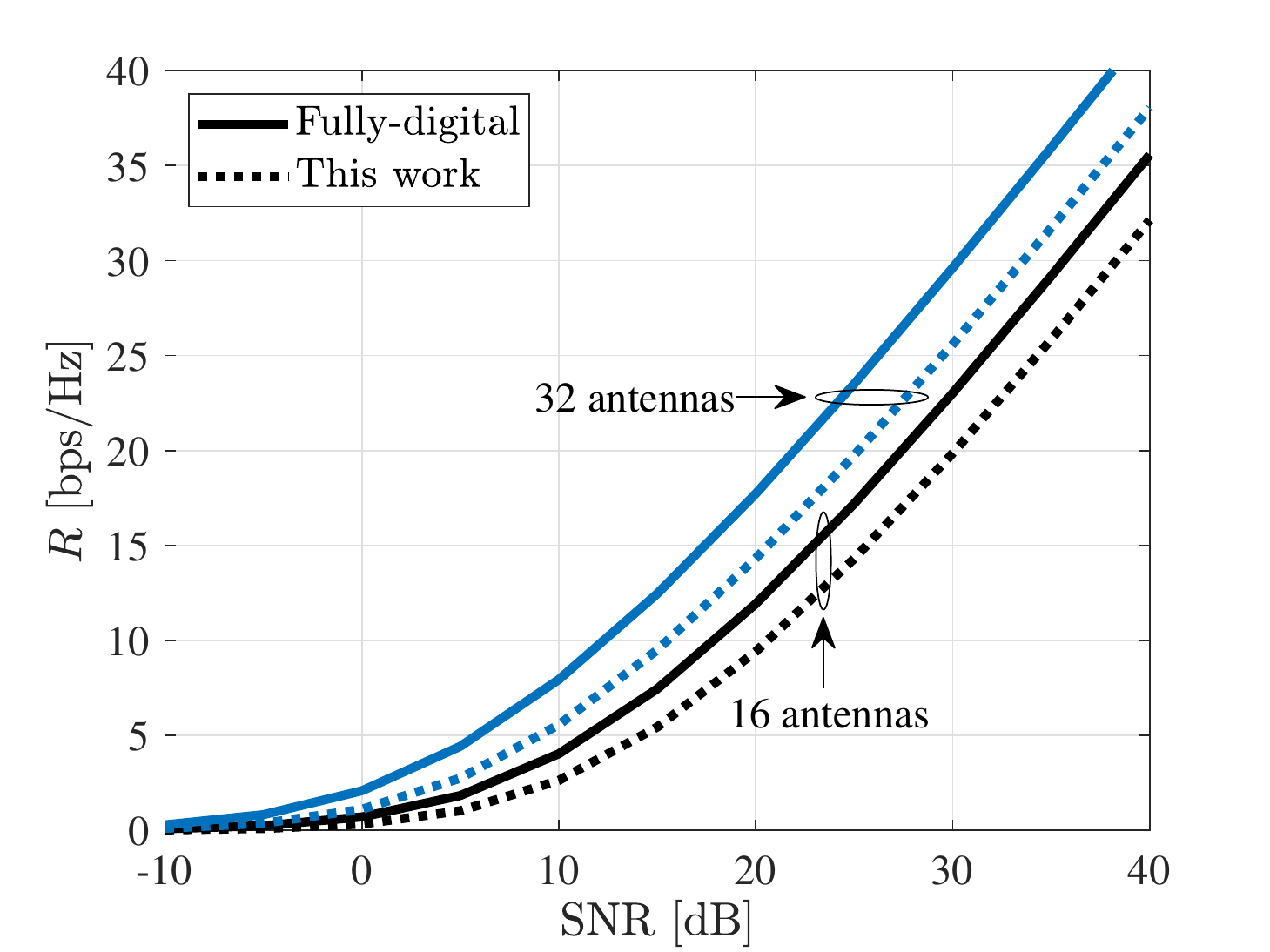}\label{fig_sumrate4users}}
\caption{Achievable sum rate $R$ as a function of $\snr = \frac{\rho}{\sigma_z^2}$ for the hybrid beamforming algorithm and ideal fully-digital BD beamforming. (a) $U=2$ users, and (b) $U=4$ users. Plots for $M_{\ap} = M_{\ue} = 16$ (black) and $M_{\ap} = M_{\ue} = 32$ (blue).}
\label{fig_sumrate}
\end{figure}

The achievable rate results for this case are shown in Fig. \ref{fig_sumrate}. As a reference, we used an ideal fully-digital beamforming architecture (one RF chain per antenna) with BD at the transmitter, eigenbeamforming at the receivers, and perfect CSI. {As we mentioned before, to the best of our knowledge, there are no other algorithms that operate under this general setting.} The beam selection procedure was made with $K_{\tx}=16$ pilot subcarriers in each case. Fig. \ref{fig_sumrate} shows the achievable sum-rate for 2 and 4 users with 16 and 32 antennas at the AP and each STA. In the asymptotically large SNR regime, the algorithm's performance is $\sim3$ dB below that of fully-digital BD regardless of the number of antennas and users. For practical SNR regimes (10 to 30 dB, see Sec. \ref{sec_raytracing}), the algorithm achieves at least 70\% of the rate with ideal BD. This performance is remarkable taking into account the hardware simplification from the fully-digital architecture (16 or 32 RF chains at the AP and each STA) to the hybrid beamforming architecture (4 RF chains at the AP and 1 at each STA). Sum-rate results with the 3-path channel model have negligible variations with respect to the single-path model and are thus omitted here.%This indicates that selection of suboptimal beamformers does not have significant impact over the performance in terms of achievable sum-rate.

%%%%%%%%%%%%%%%%%%%%%%%%%%%%%%%%%%%%%%%%%%%%%%%%%%%%%%%%%%%%%%%%%%%%%%%%%%%%%%%%%%%%%%%%%%%%%%%%%%%%%
%%%%%%%%%%%%%%%%%%%%%%%%%%%%%%%%%%%%%%%%%%%%%%%%%%%%%%%%%%%%%%%%%%%%%%%%%%%%%%%%%%%%%%%%%%%%%%%%%%%%%
%%%%%%%%%%%%%%%%%%%%%%%%%%%%%%%%%%%%%%%%%%%%%%%%%%%%%%%%%%%%%%%%%%%%%%%%%%%%%%%%%%%%%%%%%%%%%%%%%%%%%

\subsection{Ray-Tracing Validation}
\label{sec_raytracing}
%\Figure[t!](topskip=0pt, botskip=0pt, midskip=0pt)[width=0.97\columnwidth]{Figures/Scenario4.pdf}
%{Conference room scenario model in Wireless Insite\textsuperscript{\textregistered}.\label{fig_WImodel}}

\begin{figure}[!t]
\centering
\includegraphics[width=\columnwidth]{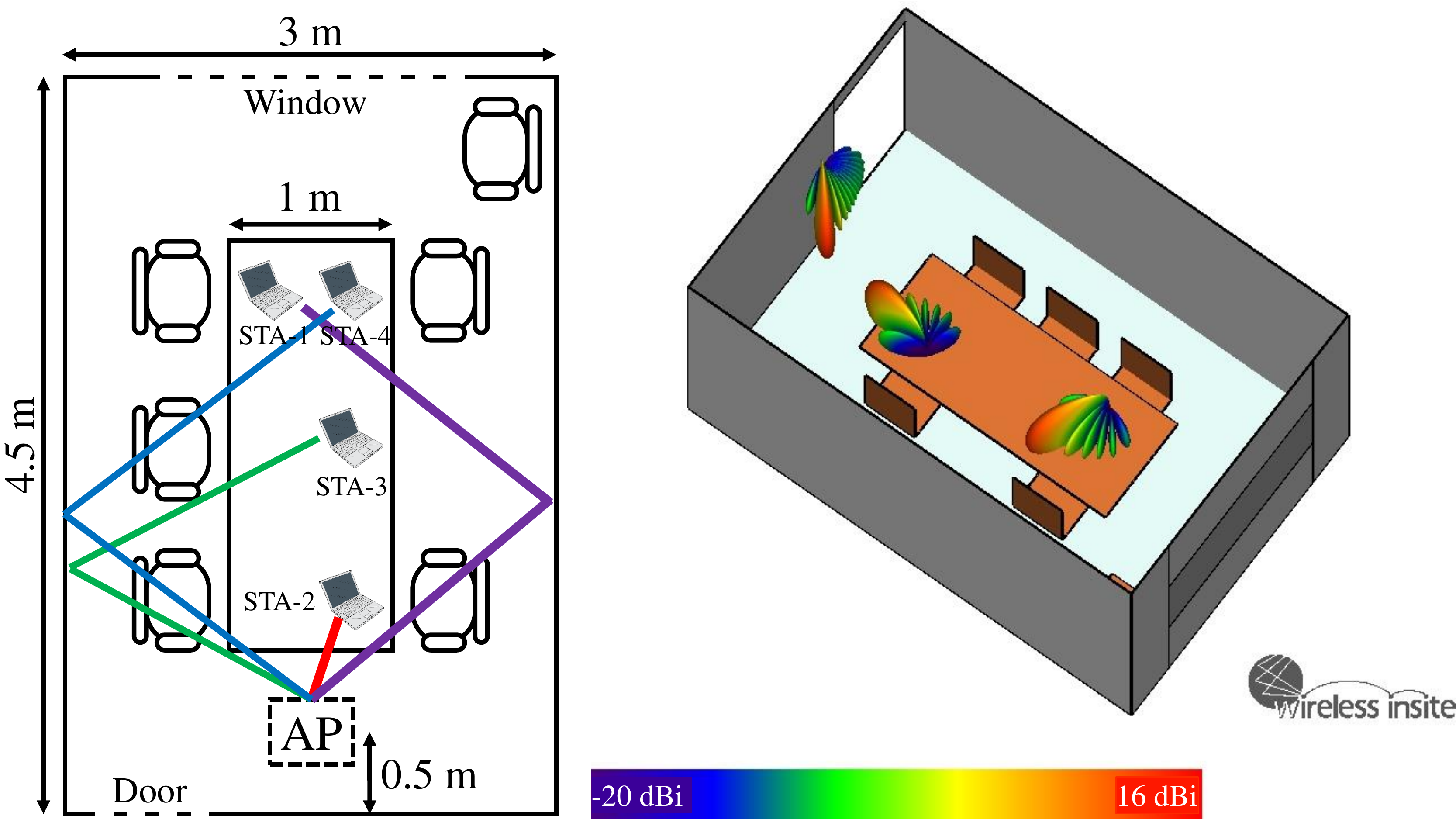}
\caption{Conference room scenario model in Wireless Insite\textsuperscript{\textregistered}.}
\label{fig_WImodel}
\end{figure}

We conducted ray-tracing simulations in scenarios specified in the IEEE 802.11ay channel models for 60 GHz WLAN \cite{maltsev2017}. Our goal is to examine the algorithm operation under real-life conditions and determine potential SNR regimes, propagation features, and algorithm accuracy. We selected the conference room evaluation scenario depicted in Fig. \ref{fig_WImodel}, where users will access wireless services such as ultra-high-definition video streaming, augmented and virtual reality, and mass-data distribution in dense hot-spots \cite{maltsev2017}. Other scenarios, such as living rooms and enterprise cubicle offices, have similar geometric characteristics and have the same use cases. We set up four STAs in the conference room, with their antenna arrays located 10 centimeters above the central table. The AP array height is 2.9 meters (10 centimeters under the ceiling), which is a typical deployment in WiFi networks. We modeled this conference room scenario using the commercial ray-tracing software Wireless Insite\textsuperscript{\textregistered} developed by REMCOM\textsuperscript{\textregistered}, as shown in Fig. \ref{fig_WImodel}. This simulator uses a diffuse scattering model to simulate scattering from surfaces at mmWave frequencies \cite{degli2007}. This approach approximates measured channel features \cite{remcom2017}. Under the diffuse scattering model, each interaction of a ray at a dielectric boundary is a source for multiple (scattered) rays, whose powers depend on a configurable angular distribution. We set the diffuse scattering parameters following the developer's recommendations to approximate measured channel impulse responses \cite{remcom2017}. The AP and STAs use planar ULAs with 16 or 32 antennas and beamforming codebooks designed with the method described in Section \ref{sec_codebook}. This type of arrays are suited for common terminals such as laptops, tablets, smartphones, TVs, and video projectors. We set up the arrays such that the beam codebook sweeps along the azimuth angle. Antenna arrays at STAs have maximum gain at $0^{\circ}$ elevation angle, while the AP is tilted $-45^{\circ}$ in the elevation angle, pointing to the middle of the conference room.

%\begin{table*}[!t]
%\caption{Algorithm Achievable Rates [bits/s/Hz] vs. BD in the Conference Room Scenario}%
%\label{tab_raytracingresults}
%\centering
%\begin{tabular}{lcccccc}
%\hline \hline
%Number of Antennas											&	User 1		&	User 2		&	User 3		&	User 4		&	Average	per User	&	Sum				\\
%\hline
%\multirow{3}{*}{\centering{$M_{\ap} = M_{\ue} = 16$}}		&	5.10 (6.83)	&				&				&				&	5.10 (6.83)			&	5.10 (6.83)		\\
%															&	3.21 (3.30)	&	5.27 (6.93)	&				&				&	4.24 (5.12)			&	8.48 (10.23)	\\
%															&	1.63 (1.68)	&	3.42 (4.71)	&	0.44 (0.95)	&	1.59 (1.59)	&	1.77 (2.23)			&	7.09 (8.93)		\\
%\hline
%\multirow{3}{*}{\centering{$M_{\ap} = 32$, $M_{\ue} = 16$}}	&	5.92 (6.87)	&				&				&				&	5.92 (6.87)			&	5.92 (6.87)		\\
%															&	3.99 (4.24)	&	7.42 (7.94)	&				&				&	5.71 (6.09)			&	11.42 (12.17)	\\
%															&	2.25 (2.45)	&	5.44 (5.95)	&	1.17 (1.71)	&	2.23 (2.41)	&	2.77 (3.13)			&	11.09 (12.52)	\\
%\hline
%\multirow{3}{*}{\centering{$M_{\ap} = M_{\ue} = 32$}}		&	6.84 (7.15)	&				&				&				&	6.84 (7.15)			&	6.84 (7.15)		\\
%															&	4.87 (5.18)	&	8.13 (8.93)	&				&				&	6.50 (7.06)			&	13.00 (14.11)	\\
%															&	3.01 (3.30)	&	6.14 (6.93)	&	1.36 (2.43)	&	2.99 (3.25)	&	3.38 (3.98)			&	13.52 (15.92)	\\
%\hline \hline
%\multicolumn{7}{l}{* All the values are rounded to 2 decimals. Values in parenthesis are the reference achievable rates with fully-digital BD.} \\
%\hline \hline
%\end{tabular}
%\end{table*}

\begin{figure}
\centering
\subfloat[]{\includegraphics[width=0.5\columnwidth]{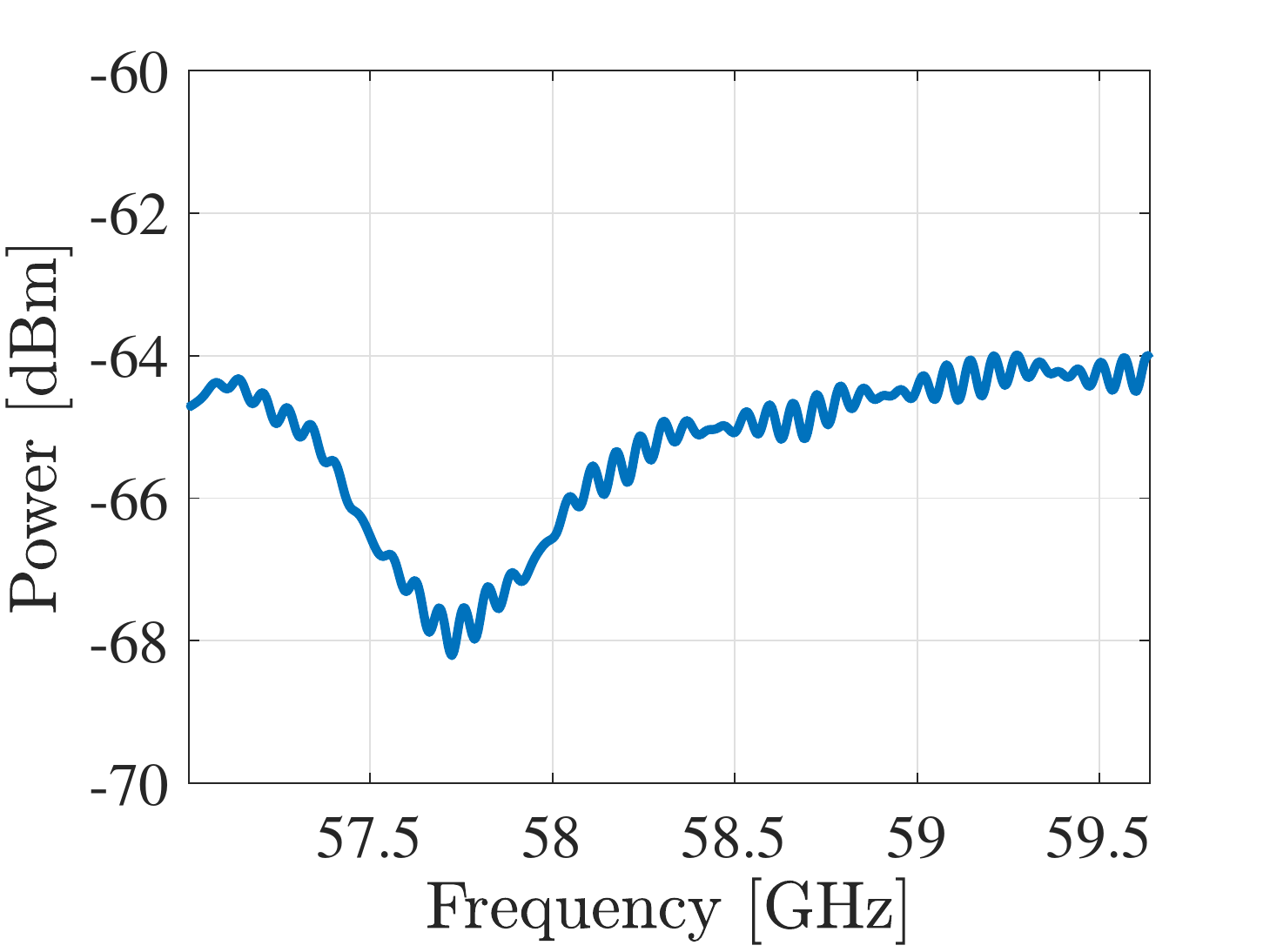}\label{fig_freqresponse}}
\subfloat[]{\includegraphics[width=0.5\columnwidth]{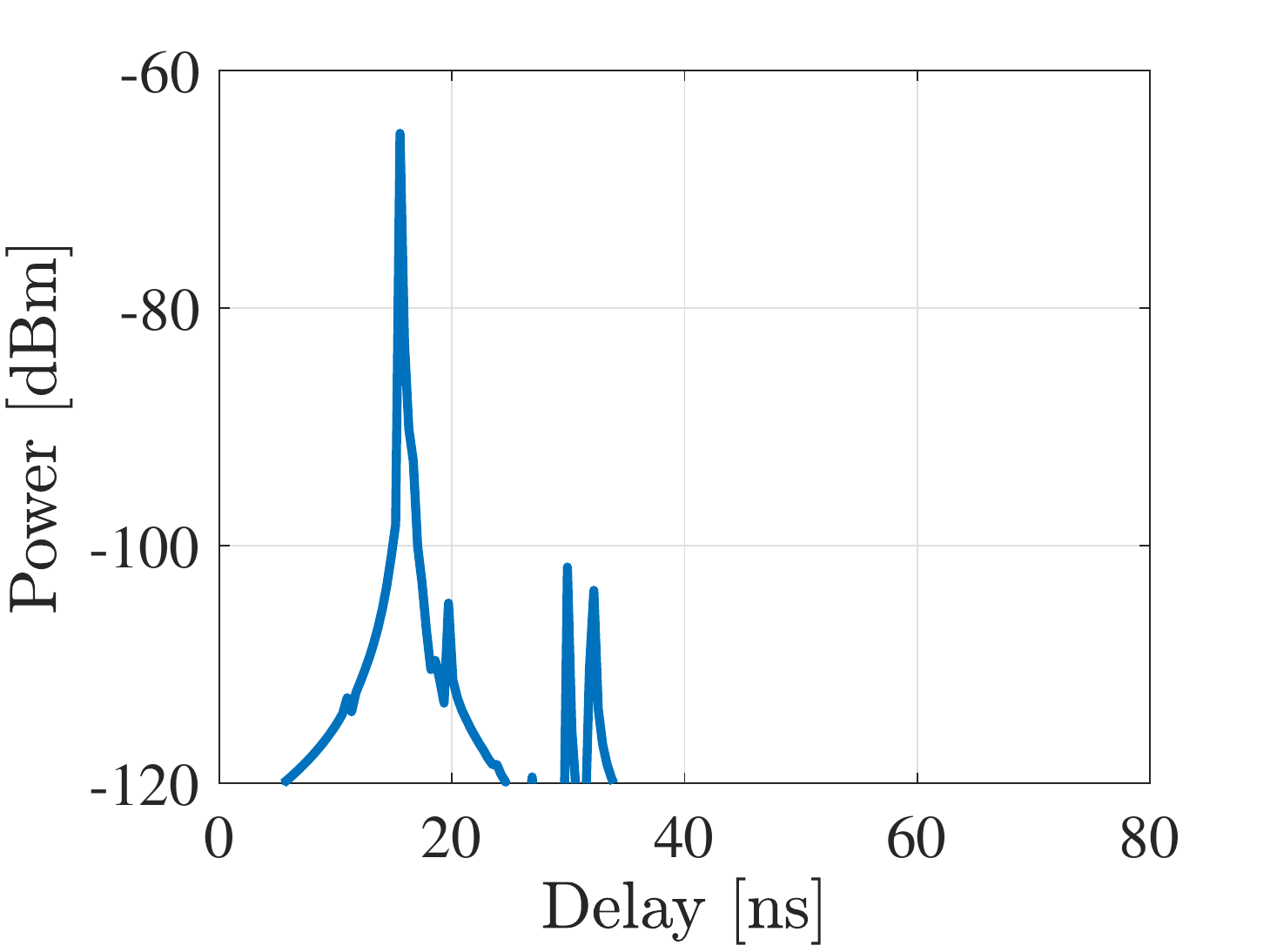}\label{fig_cir}}
\caption{(a) Channel coefficient power $| \left[ \h_{\eq,2}[k] \right]_{1,2} |^2$ from RF chain 2 to STA 2 after applying the algorithm in the scenario in Fig \ref{fig_WImodel}. (b) Equivalent power-delay profile obtained from (a).}
\label{fig_channelresponse}
\end{figure}

\begin{figure*}[!t]
\centering
\hspace*{-54pt}\includegraphics[width=2.5\columnwidth]{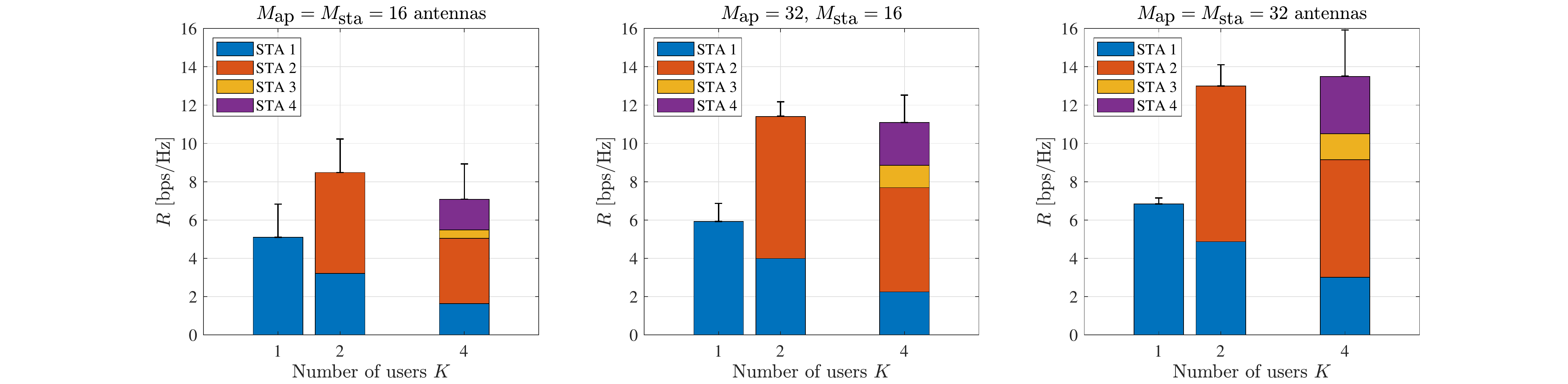}
\caption{{Algorithm achievable rates [bps/Hz] (sum and per-user rates) in the conference room ray-tracing scenario. Marks above the bars indicate the sum rate obtained with fully-digital (ideal) BD.}}
\label{fig_ratesraytracing}
\end{figure*}

Ray-tracing simulations are used to obtain frequency-dependent channel matrices for every user in the room with the parameters in Table \ref{tab_parameters}. The algorithm is then applied to this scenario, where global optima for (\ref{eq_optimbeamselection1}) are found with $\text{BSER} = 0$ after $10^3$ noise realizations. This outstanding no-error performance is due to the presence of strong line of sight or reflections and a well-oriented antenna array at the AP. Fig. \ref{fig_channelresponse} shows the frequency response between RF chain 1 and STA 1 (whose beams are paired) and also its equivalent power-delay profile obtained via inverse discrete Fourier transform. It is clear that, after beamforming, the channel impulse response is sparse. Received power varies with frequency mainly due to beam squint, but also due to weak multipath contributions.

Fig. \ref{fig_ratesraytracing} shows the achievable rates in this scenario contrasting our hybrid beamforming algorithm and the ideal fully-digital BD beamforming solution. Results are shown with a system load of 1, 2, and 4 STAs, and different antenna array sizes. A maximum spectral efficiency of 13.52 bits/s/Hz is achieved in this conference room scenario, which is sufficient to provide multi-Gbps links to all users. {The algorithm achieves sum-rates of more than 74\% of those obtained with fully-digital BD (marked above the bars), with a much simpler hardware configuration}. User 2 has the largest spectral efficiency due to its proximity to the AP, while user 3 experiments the lowest spectral efficiency due to its proximity with users 2 and 4 in the angular domain (larger power is required to suppress the interference). Users 1 and 4 have comparable performances due to their distances to the AP.

\subsection{Implementation Considerations}

Some further practical considerations for the algorithm operation are listed below:
\begin{itemize}
\item Phase shifters with $\log_2 M_{\ap}$ control bits are required at the AP and $\log_2 M_{\ue}$ bits at the STA since the number of orthogonal beamformers is equal to the number of antennas.

\item The algorithm requires the transmission of a total of $\frac{M_{\ap}}{N_{\rf}}M_{\sub} + M_{\sub} + \frac{M_{\ue}}{M_{\sub}}+1$ training sequences. Guard intervals are required between training transmissions so the phase shifters and switches can reconfigure.

\item Using the parameters in Table \ref{tab_parameters} with $M_{\ap} = M_{\ue} = 32$, $M_{\sub}=8$, and $N_{\rf}=4$, the total number of training transmissions would be 77 with a required time of approximately $3.75$ $\mu$s. Similarly, if we consider a settling time of $50$ ns to reconfigure switches and phase shifters \cite{koh2008}, the minimum time required for guard intervals would be $3.85$ $\mu$s. Thus, a total time in the order of {tens of microseconds} would be required to perform the multiuser beamforming procedure in the analyzed indoor scenarios.

\end{itemize}

Given the considerations above, the training procedure can efficiently track changes in the channel with periodic repetitions controlled by higher protocol layers.

%% file: Conclusions.tex
\section{Conclusions}

We presented a beamforming algorithm for multiuser wideband mmWave systems. The algorithm is designed to work with a fully-connected hybrid architecture at the AP consisting one antenna array, multiple RF chains, and one set of phase shifters for each RF chain. The STAs are equipped with one antenna array, one RF chain, and a switching networks with two operation modes: full-array or subarray. {The algorithm has two main contributions with respect to previous works. First, it is designed for a more general system (i.e., multiuser, wideband, with multiple hardware configurations). Second, the algorithm has reduced computational complexity and training overhead, which is achieved by leveraging the hardware configuration to avoid an explicit estimation of the channel matrix.}

{Our algorithm is based on novel beamforming codebooks with sector beams and narrow beams for hybrid and subarray hardware configurations.} The codebooks are based on the orthogonality principle of beamforming (steering) vectors of uniform arrays, which preserve the hierarchical codebook structure along the bandwidth. {We also presented system models that account for realistic antenna array effects such as beam squint, antenna coupling, and individual element radiation patterns.}

The algorithm decouples the design of analog and digital beamformers by first finding analog beamformers that maximize the received power at each STA (beam selection procedure) and then calculating a digital beamformer that eliminates inter-user interference. %For beam selection, the algorithm uses the orthogonal codebooks to transmit training signals (alternating between uplink and downlink) with the objective of finding beamformers at the AP and STA that maximize the received sum-power across subcarriers. The digital beamformer is designed using block-diagonalization to suppress inter-user interference using multiple RF chains at the AP.

We provided numerical evaluations of the algorithm in both statistical (Monte Carlo) and real-life scenarios. {First, the algorithm's performance in terms of beam selection error rate (BSER), missalignment losses, and achievable sum-rate was analyzed for three systems: single-user, multiuser with single-antenna STAs, and multiuser with one hybrid AP and subarrays at the STAs. Results show that the BSER is between 7\% and 20\% in the simulated scenarios. Missalignment losses are around 0.2 dB for 16-antenna terminals and 2.4 dB for 32-antenna terminals. In addition, errors in the analog beam selection do not significantly impact the achievable sum-rate, where the algorithm's performance is approximately 1.5 to 3 dB below of that obtained with ideal fully-digital BD.}

{The training overhead and computational complexity of our algorithm was analyzed and compared with other approaches in specific scenarios where the latter are applicable.} Our algorithm harnesses the hybrid and subarray architectures to reduce complexity, which is linear with respect to the number of antennas at the AP in contrast with the quadratic dependence of other methods. Moreover, our algorithm requires similar training overhead (or significantly less in some scenarios) as compared with channel estimation algorithms required by other beamforming methods.

{We evaluated the algorithm's performance in the conference room scenario specified in the IEEE 802.11ay mmWave WLAN standard.} We obtained channel matrices for this scenario using a commercial ray-tracing simulator that uses diffuse scattering to approximate measured mmWave channels. The algorithm achieves 13.52 bps/Hz spectral efficiency under common operating condition in this scenario when the AP and STAs have arrays with 32 antennas each. This efficiency is close to the 15.92 bps/Hz obtained with ideal fully-digital BD, which would require one RF chain for each antenna element in the arrays. Remarkably, the algorithm enables multi-Gbps connectivity to multiple users in this mmWave wideband system, with antenna and hardware limitations included.